\documentclass[preprint,review,12pt]{elsarticle}
\usepackage{titlecaps}

\AtBeginDocument{%
  \providecommand\BibTeX{{%
    \normalfont B\kern-0.5em{\scshape i\kern-0.25em b}\kern-0.8em\TeX}}}

\usepackage{geometry}
\usepackage{subcaption}
\usepackage{lineno} %linenumber
\usepackage[mathscr]{eucal} %mathscr
\usepackage{amsmath} %bmatrix
\usepackage{algorithmic} %algorithm
\usepackage[linesnumbered,lined,boxed,commentsnumbered,ruled]{algorithm2e}
\usepackage{booktabs}
\usepackage{setspace}
\usepackage{array,graphicx, multirow} % for rotate row
\newcommand{\STAB}[1]{\begin{tabular}{@{}c@{}}#1\end{tabular}}

\newcommand{\setItemSep}{\setlength\itemsep}

\usepackage{mathtools}

\usepackage[table,xcdraw, prologue, x11names]{xcolor}
\usepackage{nicefrac}
\usepackage[nocomma]{optidef}
\usepackage{tikz}
\usetikzlibrary{matrix}
\usepackage[shortlabels]{enumitem}
\RequirePackage[pdftex,
      colorlinks={true},
      linkcolor={Blue3},
      urlcolor={Blue3},
      citecolor={Blue3},
      pdfstartview={FitH},
      bookmarks={true},
            pdfauthor={Author},
            pdftitle={Title},
            pdfsubject={Subject}
            plainpages=false,
            pdfpagelabels
            ]{hyperref}
\journal{ }

\begin{document}

\title{S-DABT: Schedule and Dependency-Aware Bug Triage in Open-Source Bug Tracking Systems}

\author[1]{Hadi Jahanshahi}
\ead{hadi.jahanshahi@ryerson.ca}
\author[1]{Mucahit Cevik}
% \author[1]{Ay\c{s}e Ba\c{s}ar}

\address[1]{Data Science Lab at Ryerson University, Toronto, Canada}

\begin{abstract}
\textbf{Context}: In software engineering practice, fixing bugs in a timely manner lowers various potential costs in software maintenance. 
However, manual bug fixing scheduling can be time-consuming, cumbersome, and error-prone.

\noindent\textbf{Objective}: \textcolor{black}{In this paper, we propose the Schedule and Dependency-aware Bug Triage (S-DABT), a bug triaging method that utilizes integer programming and machine learning techniques to assign bugs to suitable developers.}

\noindent\textbf{Method}: \textcolor{black}{Unlike prior works that largely focus on a single component of the bug reports, our approach takes into account the textual data, bug fixing costs, and bug dependencies. 
We further incorporate the schedule of developers in our formulation to have a more comprehensive model for this multifaceted problem.
As a result, this complete formulation considers developers' schedules and the blocking effects of the bugs while covering the most significant aspects of the previously proposed methods.}

\noindent\textbf{Result}: \textcolor{black}{
Our numerical study on four open-source software systems, namely, {\scshape{EclipseJDT}}, {\scshape{LibreOffice}}, {\scshape{GCC}}, and {\scshape{Mozilla}}, shows that taking into account the schedules of the developers decreases the average bug fixing times. 
% Our result shows that S-DABT decreases the average fixing time of the bugs compared to its previous version, DABT.
We find that S-DABT leads to a high level of developer utilization through a fair distribution of the tasks among the developers and efficient use of the free spots in their schedules. 
Via the simulation of the issue tracking system, we also show how incorporating the schedule in the model formulation reduces the bug fixing time, improves the assignment accuracy, and utilizes the capability of each developer without much comprising in the model run times.% for generating the schedules.
}

\noindent\textbf{Conclusion}: \textcolor{black}{We find that S-DABT decreases the complexity of the bug dependency graph by prioritizing blocking bugs and effectively reduces the infeasible assignment ratio due to bug dependencies. Consequently, we recommend considering developers' schedules while automating bug triage.
}
\end{abstract}

% add your keywords

\maketitle

% \linenumbers

\section{Introduction}
In software engineering, fixing a bug as soon as possible lowers the associated costs with software development and maintenance~\citep{Kumar2017}. 
As manually determining the bug fixing decisions is time-consuming and error-prone, many researchers have been looking into the feasibility of automating each step of the procedure. 
As a bug is reported to an issue tracking system (ITS), triagers investigate its validity and determine the best course of action.
If a bug report does not include enough information for reproducibility, or if it is not relevant, it will be flagged as an invalid bug. 
However, for valid bugs, triagers may find a proper developer to assign a new bug. 
In many open software systems, developers themselves may claim a bug's possession and start fixing it. 
This decision is mainly made by checking the bug's information, e.g., its title, component, description and severity. 
\textcolor{black}{
While claiming the possession of a bug, a developer considers their available schedule and currently assigned bugs. 
Hence, both the bug's attributes and the developer's competency and free schedule are crucial in a proper bug assignment.
}

Previous studies mainly concentrate on the importance of textual information in the bug triaging process~\citep{anvik2006should, lee2017applying,mani2019deeptriage}. 
Considering the assignment task as a classical classification problem, some researchers explored the effect of considering different bug report features -- e.g., title, description, and tag -- on assigning a proper developer.
However, such approaches fail to address the issue of longer fixing times of some bugs that are automatically assigned\textcolor{black}{, e.g., due to the busy schedule of the assigned developer}. 
Other works also considered the combination of bug-fixing time (cost) and selecting the proper developer (accuracy)~\citep{kashiwa2020, park2011costriage}.
They use a regulatory term/parameter and suggest that their approach reduces the fixing time without compromising the accuracy. 
However, the algorithms still suffer from task concentration, i.e., assigning an unmanageable number of bugs to the highly expert developers. 
A recent study by~\citet{kashiwa2020} proposes setting an upper limit on the number of tasks that each developer can address given a predefined period of time. 
Their method alleviated the problem of imbalanced bug distribution and reduced the number of overdue bugs.
However, in their objective function, they do not consider the bug fixing time. 
Also, similar to many other previous works, their model does not consider the significant effect of blocking bugs in the prioritization task~\citep{akbarinasaji2018partially}. 

\textcolor{black}{\citet{jahanshahi2021dabt} proposed a Dependency-aware Bug Triaging method (DABT) that considers the bug dependency and the fixing time during the bug triage process. 
DABT employs machine learning algorithms and integer programming models to determine the suitable developers given the bug fixing times and dependencies.} 
\textcolor{black}{Nevertheless, the issue with developers' schedules and the ability to fix multiple bugs simultaneously is yet to be addressed. 
We propose Schedule and Dependency-Aware Bug Triage (S-DABT) as an extension of DABT by considering the developers' schedules, i.e., their working days, assigned bugs, holidays, and capacities for fixing bugs simultaneously. 
We also further expand dependency-awareness considerations by taking into account the dependency of the bugs assigned but not solved yet. 
% Our algorithm is called Schedule and Dependency-Aware Bug Triage (S-DABT).
}  

\paragraph{\textcolor{black}{Research Questions}} We propose S-DABT that assigns the bug to the most appropriate developer while considering the workload of available developers and the dependency between the bugs. 
In particular, we explore the following research questions.

\begin{enumerate}[start=1,label={(\bfseries RQ\arabic*)}, leftmargin=*]
    \item \textbf{How can S-DABT prevent task concentration on developers?}
    \item \textbf{How can S-DABT reduce the fixing time and the overdue bugs at the same time?}
    \item \textbf{How can S-DABT help developers postpone the blocked bugs?}
    \item \textbf{\textcolor{black}{How can S-DABT utilize the capacity of the developers compared to other baselines?}}
    \item \textbf{\textcolor{black}{How can S-DABT enhance the performance of its predecessor, DABT?}}
\end{enumerate}

\paragraph{\textcolor{black}{Extensions from Previous Work}}
\textcolor{black}{
% We extend our previous paper ``DABT: A Dependency-aware Bug Triaging Method~\citep{jahanshahi2021dabt}'' in the following ways. 
A preliminary version of this study appeared as a conference proceeding, which proposed DABT as a bug triaging method~\citep{jahanshahi2021dabt}.
The important extensions made in the current paper include the following:
\begin{itemize}
    \item We revise the model formulation such that each developer has its own simultaneous task capacity $j$. This capacity indicates the number of simultaneous bug fixes a developer can have. It is estimated based on the history of the ITS. 
    \item We define an IP model for the first time that considers developers' availability while assigning bugs. Moreover, we embed the dependency of the bugs that are assigned but not solved yet. Those bugs may delay the fixing of the blocked bugs until they get fixed.
    \item \textcolor{black}{We incorporate a new dataset, GNU Compiler Collection ({\scshape{GCC}}), to further examine the generalizability of the model.}
    \item We discuss developers' utilization and effective use of all available developers. It is reported as a new research question in the extended version. Moreover, in the online supplement\footnote{The online supplement can be found on the \href{https://github.com/HadiJahanshahi/SDABT/}{GitHub repository} of the paper.}, we explore the possible enhancement to the classification model.
\end{itemize}
}

\paragraph{\textcolor{black}{Paper Organization}}

We organized the rest of the paper as follows. Section~\ref{sec:background} briefly discusses the background of relevant methods used on our proposed bug triage algorithm, which is followed by a motivating example in Section~\ref{sec:motivating_example}.
Section~\ref{sec:research-methodology} describes the S-DABT methodology and bug report datasets used in our analysis. Section~\ref{sec:results} investigates the research questions while highlighting the importance of taking into account the bug dependency and developers' schedules. 
Finally, Section~\ref{sec:threats} discusses the study limitations and threats to validity, and Section~\ref{sec:conclusion} concludes the paper by summarizing our work and elaborating on future research directions.

%%%%%%%%%%%%%%%%%%%%%%%%%%%%%%%%%%%%%%%%%%%%%%%%%%%%%%%%%%%%%%%%%%%%%
\section{Background}\label{sec:background}
%%%%%%%%%%%%%%%%%%%%%%%%%%%%%%%%%%%%%%%%%%%%%%%%%%%%%%%%%%%%%%%%%%%%%
In this section, we briefly discuss the models employed in our proposed approach.
Specifically, we adopt Latent Dirichlet allocation (LDA) for topic modeling over bug descriptions and Integer Programming (IP) for bug triaging. 
% IP models help us construct the expected objective while imposing the desired constraints. 
Note that, to identify the suitability for the developer assignment task, we use Support Vector Machines (SVM), details of which were skipped as it is a widely used model in the domain.
% we also use Support Vector Machines (SVM) for the developer assignment task;
% however, as it is a widely used model in the domain, we do not explain them in this section.

\subsection{LDA}
LDA is a probabilistic topic modeling method~\citep{blei2003lDA}. 
It is an unsupervised learning approach that can identify the topics in given documents or corpus based on their word clusters and frequencies. 
LDA assumes that the document is a mixture of topics, and each word is attributed to one of the topics. 
In the process of bug triaging, LDA is typically used to identify the bug type given a bug report, where the latter refers to documents, and the former refers to a topic. 
That is, we extract bug description and summary from a bug report and create a bag of words (BOW) after removing the stop words. 
% It is essentially the word frequency in the bug report. 
Then, given the BOW of each bug report, LDA arbitrarily assigns each word of the report to one of the topics (i.e., bug types). 
Next, it iteratively reassigns each word, assuming that other words are accurately allocated. 
Accordingly, it computes the two conditional probabilities, the proportion of words in document $d$ that are assigned to topic $t$ --i.e., $p(\text{topic}_t|\text{document}_d)$-- and proportion of assignments to topic $t$ over all documents that come from this word as $p(\text{word}_w| \text{topic}_t)$. 
The product of these conditional probabilities gives a probability distribution based on which a new topic is assigned.

\subsection{IP Modeling}
Integer programming is a mathematical modeling framework for optimization problems where certain decision variables need to be integer-valued.
%It has a wide range of applications in various domains including healthcare, energy and manufacturing \citep{chen2010applied}.
Thanks to major advances in integer programming solution methodologies and their integration into commercial and noncommercial solvers, it has become an increasingly popular approach for various application problems over time.
Integer programming is particularly suited for scheduling problems, which typically require various integer and binary decisions and a long list of constraints to be satisfied. 
For instance, \citet{sung2016optimal} formulated an IP model for the problem of resource-constrained triage in a mass casualty incident, where the priority of the patients is identified for deploying limited emergency medical service resources so that the maximum number of patients benefit from the response efforts. 
% Their model aims to maximize the number of expected survivals and determines the transportation order of the victims as well as the destination hospitals based on the number of victims and their triage categories.

IP modeling can be similarly employed for the bug triaging problem, which involves identifying the priority of the bugs and the most suitable developers to assign those bugs. 
That is, through an appropriate mathematical model, various problem constraints \textcolor{black}{(e.g., blocking bugs and schedule availability for the developers)} can be handled while optimizing over a particular objective, such as maximizing the match between the bug descriptions and the developers. 
This way, the bug triaging can be automated to a certain extent while the overall efficiency of the process is enhanced.

% An IP model is an algebraic representation consisting of integer decision variables. 
% Integer programming is an optimization problem with a maximization or minimization objective function. 
% The knapsack problem is a specific case of IP modeling, where we have a knapsack with a limited capacity $c$ and some items. 
% Each item has its associated weight and value. The objective is to fill the knapsack with items such that it does not exceed its capacity while we include the most valuable objects in it. It can be mathematically expressed as
% \begin{align*}
% \text{maximize} & \sum_{i=1}^{n}v_{i}x_{i}\\
% \text{subject to} & \sum _{i=1}^{n}w_{i}x_{i}\leq W \\
% & x_i \in \{0,1\}
% \end{align*}
% where $x_{i}$ is the number of item $i$ with the weight of $w_i$ and the value of $v_i$. The knapsack has a maximum weight capacity of $W$. In our model, we adopt a multiple-knapsack problem with multiple capacity constraints. Hence, our problem aims to maximize the total value of the items in all knapsacks. 
% \\

\subsection{Overview of the Existing Methods}\label{sec:existing-methods}
In a typical issue tracking system, after a bug is validated, the first step is to assign it to an appropriate developer. 
Accordingly, the bug assignment is usually considered a critical task which prompted many researchers to work towards its automation. 
The proposed methods for this problem can be categorized as content-based recommendation (CBR), cost-aware bug triaging (CosTriage), release-aware bug triaging (RABT), \textcolor{black}{and dependency aware bug triage (DABT)}. 
We compare our proposed approach with the representative methods from these categories.

\subsubsection{Content-based Recommendation}
CBR approach aims to assign the most appropriate developer to the incoming bug through analyzing its content. 
\citet{anvik2006should} used machine learning techniques to build a semi-automated bug triager. 
They trained a multi-class classifier on the bug history by using SVM, Naive Bayes, and C4.5 trees, where bug title and description were converted into a feature vector as the input data, and the assigned developers were taken as the labels. 
The resulting classifier analyzes the textual contents of a given report and estimates each developer’s score for the new bug by extracting its feature vector and applying the classifier. 
% For classification, they used SVM, Naive Bayes, and C4.5 trees. 
Therefore, it can recommend suitable developers for any newly-reported bugs. 
As they reported SVM to show the best performance, we use it as the baseline classifier in our study. 
Nonetheless, some CBR studies followed the same concepts by training deep learning algorithms~\citep{lee2017applying, mani2019deeptriage, Zaidi2020}. 
\textcolor{black}{As these studies revolve around the same idea while reporting modest accuracy improvements, we consider the most commonly used approach in our analysis. 
We further examine the possible extensions to improve accuracy of the SVM classifier.}

\subsubsection{Cost-aware Recommendation}
CosTriage considers both the accuracy and the cost of an assignment.
\citet{park2011costriage} proposed a bug-triage method combining an existing CBR with a collaborative filtering recommender (CF). 
They model the developer profile that denotes their estimated cost for fixing a particular bug type. 
To create these developer profiles, they quantified their profile values as the average bug fixing time per bug type, where bug types are determined by applying the LDA to bug summary and description. 
Next, they estimated the suitability of each developer using CBR. 
They trained SVM on the textual information of the bugs and tested it for new bugs in the system. 
Finally, by combining the cost and the CBR predictions, a bug is assigned to a developer with the highest score.

\subsubsection{Release-aware Recommendation}
\citet{kashiwa2020} proposed the RABT method that primarily focuses on increasing the number of bugs resolved by the following release. 
RABT considers the time available before the next release and simultaneously keeps track of the bug fixing load on a developer. 
They formulated bug triage as a multiple knapsack problem to optimize the assignment task. 
In the standard knapsack problem, for a given set of items, each with a weight and a value, the objective is to determine the number of each items to include in a collection (i.e., knapsack) so that the total weight is less than or equal to a given limit (i.e., knapsack capacity), and the total value is as large as possible.
Similarly, the problem of assigning bugs to the developers can be considered as a variant of the knapsack problem.
That is, RABT pairs the best possible combination of bugs and developers to maximize the bug-fixing efficiency given a time limit. 
They used SVM and LDA to compute the preferences and costs for each bug and developer, respectively. 
LDA, similar to CosTriage, categorizes the bug and calculates the average time taken for each developer to fix bugs in each category. 
For a new bug, SVM computes developers' preferences, whereas LDA calculates their expected fixing time. 
Then, RABT determines the available time slot $T_t^d$ for developer $d$ based on their bugs at hand and the project horizon $\mathcal{L}$ (predetermined for each developer).
Ultimately, bugs are assigned to the most suitable developers such that their preferences are maximized, while their fixing cost does not exceed their available time slots.
% Ultimately, bugs are assigned to the most suitable developers such that their preferences are maximized, while their fixing cost does not exceed their available time slots.

% \subsubsection{\textcolor{black}{Dependency-aware Recommendation}}
% \textcolor{black}{\citet{jahanshahi2021dabt} proposed dependency-aware bug triage (DABT) method, which extended the RABT by incorporating the bug dependency considerations into their mathematical optimization framework. 
% They noted that blocking bugs defer solving blocked bugs until they are fixed. 
% Accordingly, they added a constraint to their IP model to consider bug dependency of unassigned bugs. 
% They also modified the objective function of RABT in order to optimize over both suitability and bug fixing costs.}
 
% \subsubsection{Problems with Existing Methods}
\subsubsection{Research Gaps}
The existing methods discussed above have certain limitations. 
While CBR~\citep{anvik2006should} is reportedly a highly accurate approach, it is prone to over-specialization, recommending only bugs similar to what a developer has solved before. 
Thus, it concentrates the task on some experienced developers and slowing down the bug fixing process. 
In addition, it only considers the accuracy as the performance metric, ignoring all other parameters such as bug-fixing time, the developer’s interest, and expertise. 
The CosTriage \citep{park2011costriage} method addresses the drawbacks of CBR and aims to optimize both the accuracy and bug fixing cost. 
It estimates the bug fixing time using the LDA and overcomes its sparseness through a hybrid approach and collaborative filtering recommender. 
As a result, CosTriage improves the bug fixing time without substantial degradation of accuracy.
However, both methods disregard the number of bugs a developer can fix in a given time frame. 
They may assign more bugs to experienced developers than they can address in the available time. 
Moreover, they do not consider the number of bugs that a developer can fix before the next release. 

\citet{kashiwa2020} focus on increasing the number of fixed bugs before the next release by setting an upper limit on the available time for each developer. 
Accordingly, RABT mitigates the task concentration, assigns a more achievable number of bugs that a developer can fix in a given time, and reduces the bug fixing time. 
Also, the order of assigned bugs impacts the number of bugs fixed by the release. 
If the model initially triages a time-consuming bug, it can decrease the number of bugs fixed by the next release. 
RABT also prioritizes the developers with fewer tasks as they are available to handle new bugs. 
However, it reduces the accuracy of bug assignments. 
Besides, setting the proper upper limit can be challenging. 
If the model is desired to focus on the number of bugs fixed before the next release, it should determine a dynamic threshold on developers' available times. 
Practically, by setting a constant value, the model does not become linked to the release dates. 

%%%%%%%%%%%%%%%%%%%%%%%%%%%%%%%%%%%%%%%%%%%%%%%%%%%%%%%%%%%%%%%%%%%%%%%%%
\section{\textcolor{black}{Motivating Example}}\label{sec:motivating_example}
%%%%%%%%%%%%%%%%%%%%%%%%%%%%%%%%%%%%%%%%%%%%%%%%%%%%%%%%%%%%%%%%%%%%%%%%%
\textcolor{black}{
A reported bug possesses different attributes, some of which are documented at the date of reporting while others are determined later in the fixing phase. 
With regards to a triager assigning a bug to a developer or a developer claiming its possession, the developer's schedule is typically as important as the suitability of the bug based on its attributes. 
Therefore, a proper bug triage should consider the currently assigned bugs to the developer, availability in the developer's schedule within the planning horizon, and the suitability of the new bug based on its characteristics.
}

\subsection{\textcolor{black}{Developers' Schedules}}
\begin{figure}[!ht]
    \centering
    \includegraphics[width=.9\linewidth]{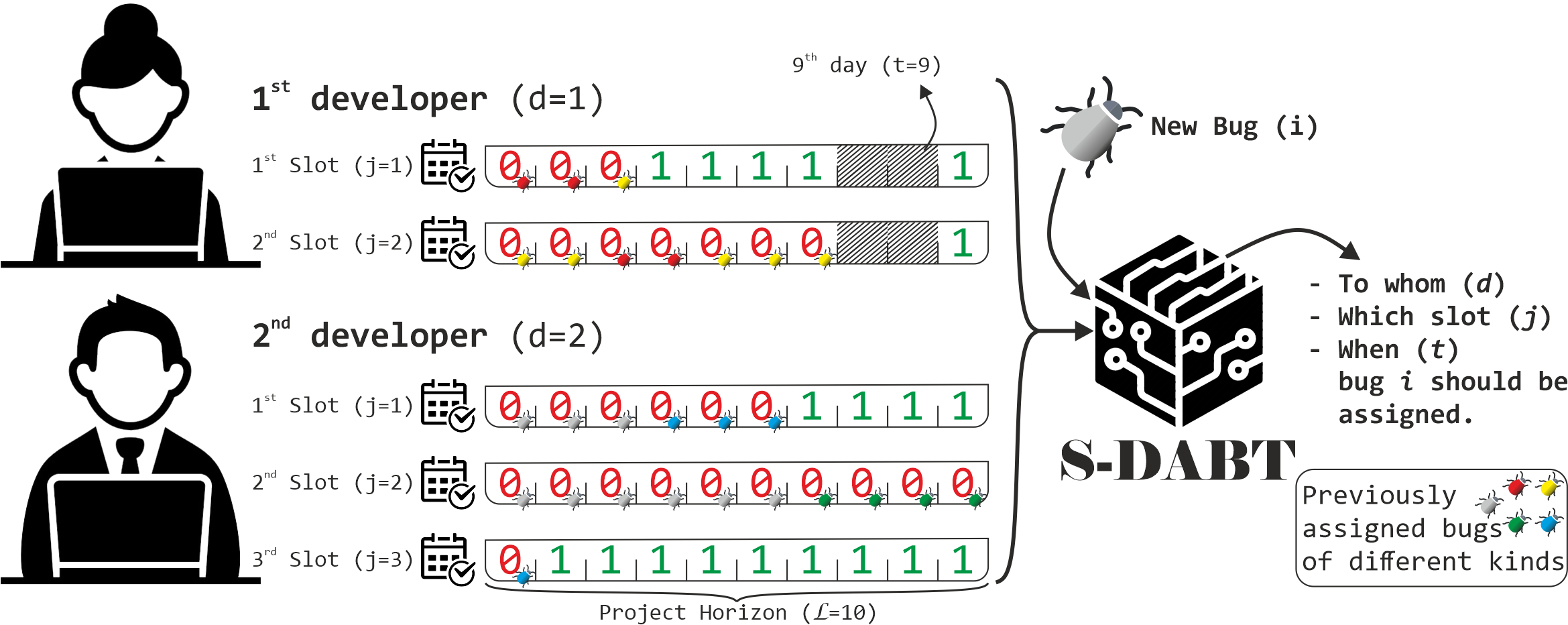}
    \caption{\textcolor{black}{Developers' schedule used for bug assignment (The bugs' color indicates their type/component).}}
    \label{fig:developer-schedule}
\end{figure}

\textcolor{black}{
To the best of our knowledge, this is the first work that takes into account the schedules of the developers while assigning a bug. 
Figure~\ref{fig:developer-schedule} shows a typical example of two developers' schedules and a new arriving bug. 
Based on the bug fixing history of each developer, we can see that the first developer can work on two bugs simultaneously, whereas the second developer may work on three bugs at a time. 
Moreover, the schedule of each developer can be estimated for the upcoming days according to the approximate fixing time of the currently assigned bugs. 
Here, ``0'' means that the schedule is occupied by an assigned bug, whereas ``1'' represents the availabilities of a developer. 
Thus, the binary parameter $T_{jt}^d$ denotes whether slot $j$ of developer $d$ is available at day $t$. 
For instance, the first slot of the second developer is preoccupied for the third day (i.e., $T_{13}^2 = 0$). 
Similarly, while assigning a bug, we may exclude the days for which a developer is not available, e.g., days 8 and 9 of developer 1.
}

\textcolor{black}{
Assume a new bug $i$ comes to the system, and a triage model is employed to automate the bug assignment process.
It first checks the suitability of the bug for each developer based on their previous experience in handling similar bugs. 
The fixing time of a bug may vary from one developer to another~\citep{kashiwa2020, park2011costriage}. 
Accordingly, the model should consider how suitable the bug is for each developer, how long the bug may take to be fixed by each developer, and how the availability of the developers will be during the planning horizon. 
For instance, assume that bug $i$ in Figure~\ref{fig:developer-schedule} takes six days for developer $d_1$, and three days for developer $d_2$ to be fixed.
% since she does not have previous experience with the bugs of the same component. 
Then, the model cannot assign the bug $i$ to developer $d_1$ as there are no six-day-long availability in her schedule. 
On the other hand, since the third slot of developer $d_2$ is almost free, the model may assign the bug $i$ to this developer.}

\subsection{\textcolor{black}{Bug Dependency}}
\textcolor{black}{
The bug dependency information can play an important role during the bug assignment process~\citep{akbarinasaji2018partially}. 
Blocking bugs are typically regarded as severe bugs that prevent some other bugs from being fixed~\citep{Hao2020}. 
Not only do they delay fixing blocked bugs, but also their fixing times are usually longer than non-blocking bugs~\citep{ValdiviaGarcia2018}. 
\citet{ValdiviaGarcia2018} noted that, in terms of cohesion, coupling complexity, and size, files affected by blocking bugs are worse than the ones affected by non-blocking issues. 
Therefore, identifying and resolving them as early as possible is crucial in issue-tracking systems.
% \citet{jahanshahi2021dabt} defined an IP model for determining bug assignment decisions, which incorporates the dependency relations between the bugs. 
% However, their model does not consider the dependency between in-progress blocking bugs and the new bugs, which we aim to address in our study.
% Such uncertainty is yet to be addressed.
}

\begin{figure}[!ht]
\centering
    \begin{tikzpicture}[b/.style={circle,draw,execute at begin node={$b_{#1}$},
        alias=b-#1,label={[rectangle,draw=none,overlay,alias=l-#1]right:{$[s_{#1}^d,c_{#1}^d]$}}}]
    \node[matrix of nodes,column sep=1em,row sep=2em]{
     & & |[b=1]|& & |[b=2]| & & &|[b=9]|\\
     &  |[b=3]|& & |[b=4]| &  & &|[b=7]| &\\
       |[b=5]|& & |[b=6]| &  & & &|[b=8]| &\\
    };
    \path[-stealth] foreach \X/\Y in {1/3,3/5,3/6,1/4,2/4,7/8} {(b-\X) edge (b-\Y)};
    \path (l-7.east); %<- for the bounding box
    \end{tikzpicture}
    \caption{A typical BDG ($b_i$: bug $i$, $s_i^d$: suitability of bug $i$ for developer $d$, $c_i^d$: fixing time of bug $i$ for developer $d$)}
    \label{fig:BDG}
\end{figure}
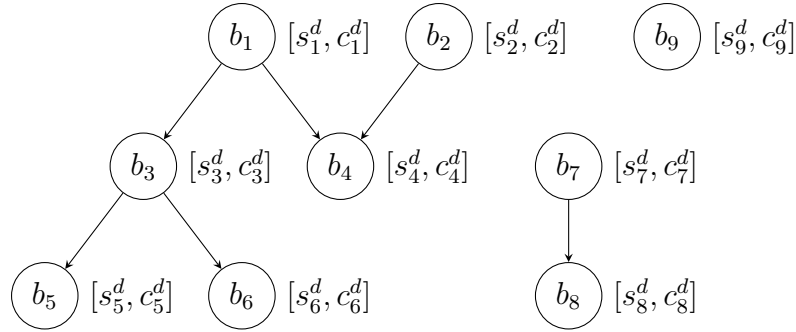
    
\textcolor{black}{Figure~\ref{fig:BDG} shows the dependencies among various bugs together with bug attributes such as fixing cost $c$ and developers' suitability $s$.
For example, regarding bug dependency, $b_3$ is the blocking bug for $b_5$ and $b_6$. 
Thus, fixing those two bugs is constrained by fixing the parent node (i.e., the blocking bug). 
Moreover, the resolution of $b_3$ should be postponed until its blocking bug $b_1$ is resolved. 
Such dependencies may raise maintenance costs, lower overall quality, and cause the software systems' release to be delayed. 
We note that solo bugs are dominant and omnipresent in such dependency graphs (e.g., see $b_9$), neither blocking nor being blocked by other bugs. 
Accordingly, a proper bug assignment model should consider the bug dependency while assigning the bugs to each developer. 
It should prioritize the ones blocking other bugs to reduce their negative impact in the long run. 
}

\section{Methodology}\label{sec:research-methodology}
%%%%%%%%%%%%%%%%%%%%%%%%%%%%%%%%%%%%%%%%%%%%%%%%%%%%%%%%%%%%%%%%%%%%%
We examine the bug evolution in software repositories and its effect on the triaging process. 
We consider four open software systems, namely, {\scshape{EclipseJDT}}, {\scshape{LibreOffice}}, {\scshape{GCC}}, and {\scshape{Mozilla}}. 
We extract reported bugs' information from their Issue Tracking System, covering a decade from January 2010 to December 2019. 
In the triaging process, we consider both textual information of the reported bugs and their dependencies. 
We construct a bug dependency graph (BDG), where the reported bugs are its nodes, and the blocking information, i.e., ``depends on'' and ``blocks'' forms its arcs. 
A BDG is a directed acyclic graph (DAG) in which bugs cannot block other bugs while they are simultaneously blocked by the same bugs in a loop (see Figure~\ref{fig:BDG}). 
Bug dependencies are usually identified during the fixing process; hence, we assume the BDG is constantly evolving by the changes in its nodes -- i.e., fixing a bug or introducing a new bug -- and its arcs -- i.e., finding a new dependency or removing an existing dependency.

As we discussed in Section~\ref{sec:existing-methods}, the textual information and the fixing time of the bugs are of importance in automating the bug triage processes. 
Previous studies focusing on CosTriage-based methods claim that merely looking for the textual information may increase the accuracy; however, such an approach leads to longer bug fixing times. 
Therefore, they define a control parameter, $\alpha$, to make a trade-off between accuracy and the fixing cost/time. 
Still, both CBR and CosTriage approaches ignore the impact of the imbalanced distribution of the fixing tasks among developers. 
The release-aware bug triaging method enriches the triaging process with a predefined constraint on developers' workload to minimize overdue bugs. 
% \textcolor{black}{
% In our previous work, we incorporate the bug dependency relations that are represented as BDGs into the bug assignment decisions~\citep{jahanshahi2021dabt}. 
% We formulate an integer programming model and include a constraint set to account for the bugs that are not fixable due to various dependencies to other bugs. 
% However, the schedules of the developers are yet to be incorporated while determining the bug assignment decisions, which we set out to achieve below. 
% We discussed the importance of developers' schedule in the bug triage problem in Section~\ref{sec:motivating_example}.
% }

% \subsection{Model's Assumptions}
\subsection{Assumptions}
We make the following assumptions in order to formulate an integer programming model for the problem:
\begin{itemize}\setItemSep{0.5em}
    \item Similar to previous works\textcolor{black}{~\citep{anvik2006should, park2016cost, kashiwa2020}}, we only consider \textit{active} developers in the ITS, and exclude inactive developers as we do not have sufficient information on them. 
    Considering interquartile range (IQR) as a measure of central distribution, we define \textit{active} developers as the ones whose bug fix number is higher than the IQR of bug fix numbers of all developers. \textcolor{black}{We acknowledge that the definition of \textit{active} developers is time-dependant in the agile software industry. That is, even an \textit{active} developer may leave the job at some point. Accordingly, the list of \textit{active} developers requires constant updates in practice~\citep{Zhang2020}.}
    
    \item We consider a fixed number of developers, $d \in \{d_1, d_2, \dots, d_D\}$ working on resolving bugs \textcolor{black}{by taking into account their schedule.} 
    We obtain the list of developers using predefined conditions on \textit{active} developers. Specifically, we take $D$ as 16, 48, 47, and 121 for {\scshape{EclipseJDT}}, {\scshape{LibreOffice}}, {\scshape{GCC}} and {\scshape{Mozilla}}, respectively. 
    
    \item We do not incorporate the severity and priority information of the bugs into the bug assignment decisions as they were reported to be unreliable and subjective~\citep{akbarinasaji2018partially, jahanshahi2020Wayback}. 
    
    \item Bug $i$ has fixing time of $c_i^d$ if it is solved by developer $d$. 
    Also, $c_i = \{c_i^{d_1}, c_i^{d_2}, \hdots, c_i^{d_D}\}$ is used to denote the set of cost values of bug $i$ for all the active developers.
    For our analysis to be consistent and comparable with those of previous studies~\citep{kashiwa2020, park2011costriage}, we estimate the fixing time using the same approach reported in those works. 
    We apply LDA for topic modeling, use Arun's method to obtain the optimal number of categories, find the average fixing time of each developer given the category, and finally, estimate the missing values using a collaborative filtering recommender. 
    \textcolor{black}{Arun's method determine the optimal number of bug types as 8, 18, 5, 5 for {\scshape{EclipseJDT}}, {\scshape{LibreOffice}}, {\scshape{GCC}} and {\scshape{Mozilla}}, respectively.}

    \item Bug $i$ has a \textit{suitability} $s_i^d$ when assigned to developer $d$. 
    Also, $s_i = \{s_i^{d_1}, s_i^{d_2}, \hdots, s_i^{d_D}\}$ denotes the set of suitability values of bug $i$ for all developers.
    The notion of suitability implies that, in the triage process, we should assign the bugs to the most compatible developer. 
    To calculate the suitability, we train a model on textual information (i.e., bug title and description) obtained from the history of resolved bugs.
    A TF-IDF model converts the merged textual data to numeric values and makes those ready for the classifier. 
    We adopt an SVM classifier that employs the output of TF-IDF as the independent features and the developers' names as the class labels. 
    We fit the model at the end of the training phase. 
    Then, it can predict the suitability of each developer given the textual information of a new bug. 
    We adopt SVM implementation from the scikit-learn library in Python with a linear kernel and the regularization parameter $C=1000$. 
    Note that these settings are compatible with the previous works. 
    We use default values for other SVM parameters. 
    Figure~\ref{fig:suitability} shows the process of estimating $s_i^d$ values.
    \vspace{-0.2cm}
    \begin{figure}[!ht]
    \vspace{-0.1cm}
        \centerline{\includegraphics[width=.7\linewidth]{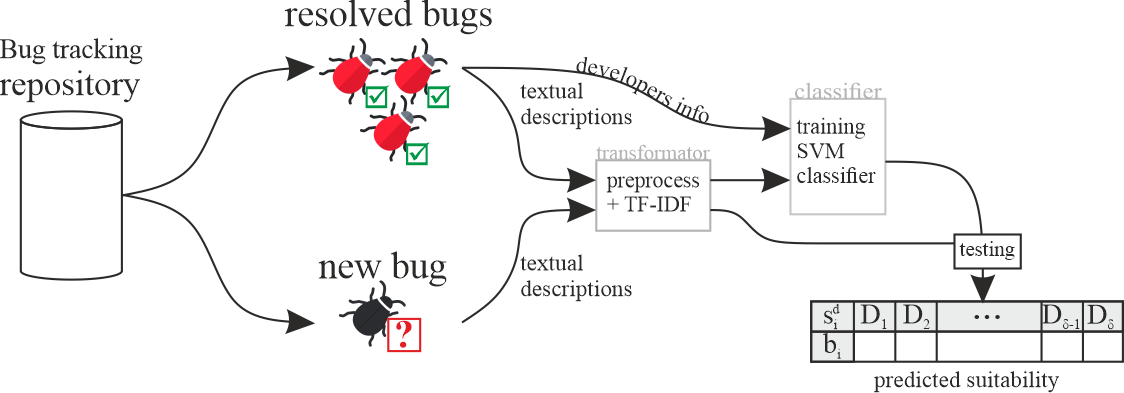}}
        \caption{Suitability estimation process~\citep{kashiwa2020}}
        \label{fig:suitability}
    \end{figure}
    
    \item Similar to \citet{kashiwa2020}, we solve the IP model and other baseline methods at the end of each testing day. 
    Accordingly, developers are assigned bugs once a day \textcolor{black}{(i.e., the interval between bug assignments is 24 hours)}.
    
    \item Each bug cannot be assigned to more than one developer at the same time.
    
    \item Bug dependency is updated at the time of finding a dependency or removing one. 
    Therefore, dependency updates are instantaneous, separate from the model runs and are done by the Wayback Machine simulator~\citep{jahanshahi2020Wayback}.
    
    \item \textcolor{black}{
    Each developer may work on a fixed number of bugs on a given day, which depends on the ``capacity'' of each developer. 
    We estimate developers' capacity according to the history of bug assignments (i.e., how many bugs have been historically assigned to each developer on a single day.). 
    We refer to this capacity parameter as \textit{schedule slots}, which is denoted by $j$.}

\end{itemize}

\subsection{\textcolor{black}{Estimating Developers' Schedules}}
\textcolor{black}{
    Estimating the developers' capacity in the OSS systems according to their previous schedule can be cumbersome since particular developers can be part-time employees or volunteer contributors. However, this is typically feasible in proprietary software projects in which the work schedules of developers are largely predetermined. In the OSS systems, there are still many employed developers whose working hours are known. Accordingly, our proposed model can be fed with the actual schedules of the recruited staff, and it can associate each bug with one of the available developers. Note that, in this work, we do not have access to the actual schedules of the developers; instead, we estimate their working hours based on the available information in the OSS repository. In practice, a scheduling system can be incorporated into the bug triage routine in which each developer may report or modify their available schedules for the upcoming week. As the assignment is done at the end of each day, they may adjust their schedules at the end of the working day to have an appropriate bug assignment for the next day.
}

\textcolor{black}{
    We assume that, based on the proficiency and experience of each developer, they may handle $j\geq 1$ bugs simultaneously. We estimate the number of slots (i.e., capacity to address multiple bugs at the same time) of each developer according to our observations in the training phase. Specifically, we track the number of assigned bugs to each developer and report the number of daily bugs they work on during the training period. Through such approximation, we can estimate the capability of a certain developer in handling multiple bugs. Using the third quartile and interquartile range of previous simultaneous tasks, we approximate the developer capacity, i.e., $j \leq \lfloor\text{Q}3 + (1.5\times\text{IQR})\rfloor$. We acknowledge that having access to the exact schedule of the developers results in more accurate parameters for our models. However, our approach to incorporating developers' schedules into the bug triaging process remains valid even with this estimated parameter set for their schedules.
}

\textcolor{black}{
    We also note that having a binary vector representation of the developers' schedules (e.g., $[1,0,0,1,0]$ indicating the developer being available only at 1\textsuperscript{st} and 4\textsuperscript{th} time steps) enables the project managers to insert the exact working hours or off days for the developers to have appropriate bug assignments. Moreover, although we apply the model on a daily basis, the granularity of the schedules may vary. The model can also accommodate other schedule breakdowns (e.g., hourly or weekly). Since we estimate the fixing times of the bugs on a daily basis, we split the developers' schedules according to their daily availability. The definition of the breakdowns can be easily adjusted based on the triagers' preferences. 
}

\subsection{Dependency-aware Bug Triage}
We first explain the DABT model that we originally proposed to incorporate bug dependency relations into bug triage decisions~\citep{jahanshahi2021dabt}. Let $x_i^d$ be a binary variable that takes value 1 if bug $i$ is solved by developer $d$, and 0 otherwise. In addition, let $T_t^d$ denote the time limit of developer $d$ at time $t$. Specifically, at time step $t$, developers' time limit is updated according to their schedule and previously assigned bugs. We define an identical upper limit $\mathcal{L}$ for all developers on their $T_t^d$ values. This upper limit is equivalent to the maximum planning horizon for a project and can change according to the project size. We set $\mathcal{L}$ to the third quartile value of the bug fixing times according to the previous study~\citep{kashiwa2020}. 
% The value of $\mathcal{L}$ for  {\scshape{EclipseJDT}}, {\scshape{LibreOffice}}, {\scshape{GCC}} and {\scshape{Mozilla}} is 26, 10.5, 15, and 18 days, respectively. Also, we use $P(i)$ to denote the list of parents of bug $i$. 
% The IP model for bug assignment can then be formulated as follows.
DABT model is formulated as follows:
\begin{maxi!}|l|
  {}{\sum_{d} \sum_i \Big(\big(\alpha \times \frac{s_i^d}{\max(s_i)}\big) + \big( (1-\alpha) \times \frac{\nicefrac{1}{c_i^d}}{\nicefrac{1}{\min(c_i)}} \big) \Big) x_i^d  \label{eq:IP-objective-old}}{}{}
  \addConstraint{x_i^d}{\leq x_p^d}{\quad \forall d, \ \  \forall i \neq \text{root}, p \in P(i) \label{eq:IP-1}}
  \addConstraint{\sum_i c_i^d x_i^d}{\leq T_t^d }{\quad\forall d }\label{eq:IP-2}
  \addConstraint{\sum_{d} x_i^d}{\leq 1}{\quad\forall i \label{eq:IP-3}}
  \addConstraint{x_i^d}{\in \{0,1\}}{\quad\forall i \label{eq:IP-4}}
 \end{maxi!}

The objective function maximizes the suitability of the solved bugs while reducing the cost (i.e., solving time). The trade-off between suitability and estimated fixing time is determined by $\alpha$. Different from \citet{kashiwa2020}'s model, we incorporate the fixing time in the objective function since ignoring the solving time might cause over-specialization. It also helps to reduce bug fixing time. Constraints~\eqref{eq:IP-1} ensure that the precedence constraints are imposed, that is, blocking bugs $x_p^d$ are solved before the blocked bug $x_i^d$. Constraints~\eqref{eq:IP-2} enforce total time limit requirements. Lastly, constraints~\eqref{eq:IP-3} guarantee that a single bug cannot be assigned to more than one developer simultaneously.

\subsection{\textcolor{black}{Schedule- and Dependency-aware Bug Triage}} \label{sec:SDABT}
\textcolor{black}{
DABT extended previous approaches by adding a bug dependency constraint and minimizing bug fixing times in the objective function. 
However, the bug dependency is only restricted to the unassigned bugs. 
Blocking bugs that are currently assigned can still impede the blocked bugs. 
Therefore, a comprehensive model should consider those blocking bugs to be able to realize the full potential of dependency-aware bug triage. 
Another salient point of bug assignment is the developers' schedule. 
Not all developers have the same capacity for solving bugs, and especially in open-source software (OSS) systems, some are more active, and some are freelancers. 
Therefore, their dedicated time to address the bugs may differ. 
Our proposed model defines developers' schedules based on their bug fixing history. 
That is, they can handle one or more bugs simultaneously, depending on their schedule availability.
}

\textcolor{black}{
We define our Schedule and Dependency-aware bug triage (S-DABT) model as an extension of DABT. 
S-DABT determines the assignment of the bugs to the developers based on developers' schedule and competency together with bugs' fixing cost and dependency. 
Let $x_{ijt}^d$ be a binary variable that takes value 1 if bug $i$ is assigned to slot $j$ of developer $d$ starting from day $t$, and 0 otherwise. 
The bug takes $c_i^d$ days to be solved and has the preference (i.e., suitability) of $s_i^d$ for this assignment. 
In addition, let $T_{jt}^d$ denote a binary parameter that represent developer $d$'s availability, which take value 1 if slot $j$ at day $t$ is free, and 0 otherwise. 
Specifically, at each time step, developers' schedule is updated according to previously assigned bugs, possible holidays, and off days. 
The number of days that we plan ahead for has the upper limit $\mathcal{L}$ for all developers (i.e., $t \in \{1, 2, \dots, \mathcal{L}\}$). 
This upper limit is equivalent to the maximum planning horizon for a project and can change according to the software project specifications and size. 
We set $\mathcal{L}$ to the third quartile value of the bug fixing times, as reported in previous studies~\citep{kashiwa2020}. 
The values of $\mathcal{L}$ for {\scshape{EclipseJDT}}, {\scshape{LibreOffice}}, {\scshape{GCC}} and {\scshape{Mozilla}} are 9, 3, 15, and 10 days, respectively. 
Also, we use $P(i)$ to denote the list of parents for bug $i$. 
The integer programming model for bug assignment can then be formulated as follows.}\\

% \begin{footnotesize}
% \begin{subequations}
% \begin{align}
% \text{maximize} \ & \sum_i \sum_d \sum_j \sum_t \Big(\big(\alpha \frac{s_i^d}{\max(s_i)}\big) + \big( (1-\alpha) \frac{\nicefrac{1}{c_i^d}}{\nicefrac{1}{\min(c_i)}} \big) \Big) x_{ijt}^d &&\\
% \nonumber \text{s.t.} \ & \big(1 - \sum_d \sum_j  \sum_t x_{ijt}^d \big)  M +  \sum_d \sum_j  \sum_t {(x_{ijt}^d  t)} & > & \sum_d \sum_j  \sum_t \big(x_{pjt}^d  (t+c_i^d-1)\big)\\ 
% \ &&&& \hspace{-4.5cm} \quad \forall i \neq \text{root} \text{ if } p \in P(i), P(i) \neq \emptyset
% \end{align}
% \end{subequations}
% \end{footnotesize}

{\footnotesize
\begin{maxi!}|l|
  {}{\sum_i \sum_d \sum_j \sum_t \Big(\big(\alpha \times \frac{s_i^d}{\max(s_i)}\big) + \big( (1-\alpha) \times \frac{\nicefrac{1}{c_i^d}}{\nicefrac{1}{\min(c_i)}} \big) \Big) x_{ijt}^d  \label{eq:IP-objective}}{}{} %%%%%%%%%% a %%%%%%%%%% 
  \addConstraint{\sum_d \sum_j \sum_t x_{ijt}^d}{\leq 1}{\hspace{-3.4cm} \forall i, \label{eq:IP-single-assignment}}  %%%%%%%%%% b %%%%%%%%%% 
  \addConstraint{x_{ijt}^d}{\leq T_{jt^{\prime}}^d }{\hspace{-3.4cm} \forall i,d,j,t , t^{\prime} \in [t, \dots, t+c_i^d-1], \label{eq:IP-slot-time-availability} } %%%%%%%%%% c %%%%%%%%%% 
  \addConstraint{x_{ijt}^d}{= 0}{\hspace{-3.4cm} \forall i,d,j,t \text{ if } t+c_i^d-1 \geq \mathcal{L}, \label{eq:IP-bug-solution-check} } %%%%%%%%%% d %%%%%%%%%% 
  \addConstraint{\sum_{i}\sum_{t^{\prime}=t-c_i^d+1}^{t} x_{ijt^{\prime}}^d}{\leq 1}{\hspace{-3.4cm} \forall d, j, t; \text{ if } T_{jt^\prime}^d = 1, \label{eq:IP-one-bug-at-time-t}} %%%%%%%%%% e %%%%%%%%%%
%   \addConstraint{\sum_{{i^\prime}\neq i}\sum_{t^{\prime}=t}^{t+c_i^d-1} x_{{i^\prime}jt^{\prime}}^d}{\leq (1-x_{ijt}^d)M}{}{\nonumber} 
%   \addConstraint{}{}{\hspace{-3.4cm} \forall i, d, j, t; \text{ if } T_{jt}^d = 1, \label{eq:IP-one-bug-at-each-slot}}  %%%%%%%%%% e %%%%%%%%%% 
  \addConstraint{\big(1 - \sum_d \sum_j  \sum_t x_{ijt}^d \big)  M +  \sum_d \sum_j  \sum_t {(x_{ijt}^d t)} }{> \sum_d \sum_j  \sum_{t^\prime} \big(x_{pj{t^\prime}}^d  (t^\prime+c_p^d-1)\big)}{\nonumber} 
  \addConstraint{}{}{\hspace{-3.4cm} \forall i \neq \text{root}, \text{ if } p \in P(i), P(i) \neq \emptyset, \label{eq:IP-dependency1}}  %%%%%%%%%% f %%%%%%%%%% 
  \addConstraint{\big(1 - \sum_d \sum_j  \sum_t x_{ijt}^d \big)  M +  \sum_d \sum_j  \sum_t {(x_{ijt}^d  t)} }{> \tau_p }{\hspace{-3.4cm} \forall i \neq \text{root}, p \in P^{\prime}(i), \text{ if } P^{\prime}(i) \neq \emptyset, \label{eq:IP-dependency-assigned-not-solved}}  %%%%%%%%%% g %%%%%%%%%% 
  \addConstraint{\sum_d \sum_j \sum_t x_{ijt}^d}{\leq \sum_d \sum_j \sum_t x_{pjt}^d}{\nonumber}
  \addConstraint{}{}{\hspace{-3.4cm} \forall i \neq \text{root}, p \in P(i), \text{ if } P(i) \neq \emptyset, \label{eq:IP-dependency3}}  %%%%%%%%%% h %%%%%%%%%% 
  \addConstraint{x_{ijt}^d}{\in \{0,1\}}{\hspace{-3.4cm} \forall i, j, t, d \label{eq:IP-binary}}  %%%%%%%%%% i %%%%%%%%%% 
\end{maxi!}
}

\textcolor{black}{The objective function of the model maximizes the suitability of the solved bugs while reducing the cost (i.e., solving time). 
The trade-off between suitability and estimated fixing time is determined by the parameter $\alpha$.
Different from \citet{kashiwa2020}'s model, we incorporate the fixing time in the objective function since ignoring the solving time might cause over-specialization. 
It also helps to reduce bug fixing time. 
Constraints~\eqref{eq:IP-single-assignment} guarantee that a single bug cannot be assigned to more than one developer simultaneously. 
Equations~\eqref{eq:IP-slot-time-availability} ensure that if bug $i$ is assigned to developer $d$, then the developer has sufficient available days in the schedule (i.e., $T_{jt^\prime}^d$ =1) until the bug is completely fixed (i.e., the end of the fixing period (from $t$ to $t+c_i^d-1$)). 
Moreover, constraints~\eqref{eq:IP-bug-solution-check} restrict assignments to developers if the fixing date exceeds the project horizon ($t+c_i^d-1 \geq \mathcal{L}$). 
On the other hand, when bug $i$ is assigned to slot $j$ of developer $d$, equations~\eqref{eq:IP-one-bug-at-time-t} ensure that no other bug can be assigned to the same slot until bug $i$ is fixed.}

\textcolor{black}{Regarding the blocking bugs, the model is required to formulate dependencies both among unassigned bugs and between unassigned and assigned bugs. Hence, we define three constraints to account for all different cases. Assume $p$ is one of the parents of bug $i$ (i.e., $p$ blocks bug $i$). Constraints~\eqref{eq:IP-dependency1} ensure that the blocked bug can only be solved after its parent is resolved. 
In that case, if bug $i$ is assigned to be solved starting at time $t$ (i.e., $x_{ijt}^d=1$), its blocking bug $p$ should have finished its fixing period ($t > t^\prime+c_p^d-1 $). 
If the bug is not yet assigned (i.e., $x_{ijt}^d=0$), the Big-$M$ parameter assures that the constraint always holds. 
We set Big-$M$ parameter to $\mathcal{L}$, which is the length of the maximum project horizon (i.e., $t^\prime+c_p^d-1$ cannot be greater than $\mathcal{L}$). 
Beside the importance of assigning a bug and its parent at the same time, there might be cases in which the parent (blocking bug) is assigned, but it is yet to be fixed. Equation~\eqref{eq:IP-dependency-assigned-not-solved} guarantees that fixing the blocking bug that is already assigned is earlier than the assignment time of the blocked bug ($i.e., \tau_p < t$). 
Lastly, constraints~\eqref{eq:IP-dependency3} condition the assignment of bug $i$ to the assignment of its blocking bug $p$. \textcolor{black}{We note that our model does not impose any constraints to assign a blocking bug to the same developer as the blocked one. That is, a set of dependent bugs can be solved by multiple developers according to the solution sequence of the blocking bugs. On the other hand, if, in a particular setting, it is desired that a set of dependent bugs need to be solved by a single developer, a corresponding set of constraints can be added to our model to satisfy this requirement. }
}

\textcolor{black}{The proposed IP formulation incorporates all the previous criteria while it improves dependency constraints and accounts for the schedules of the developers. 
Accordingly, the IP formulation proposed by~\citet{kashiwa2020}
%and ~\citet{jahanshahi2021dabt} are 
is different than ours in that it disregards the bug dependency and the schedule of the developers, the blocking effect of already assigned bugs, and the exact timing and order for the assignments. 
We provide an illustrative example of our proposed model in the online supplement\footnote{The online supplement can be found on the \href{https://github.com/HadiJahanshahi/SDABT/}{GitHub repository} of the paper.}.
}

% \subsection{Experimental setup}
\subsection{Bug Triage Mechanism}
We design a mechanism to recreate all the past events.
Unlike the previous studies that worked on the static datasets (typically stored in CSV or JSON files), we reconstruct the history of events and apply each algorithm in real-time. 
The online exploration enables us to examine the outcome of bug assignments given other elements in the system.
Therefore, we build a pipeline of the past events and replace the assignment task with our proposed algorithm. 
The flow and time of the bug reports, bug reopening, and bug dependency discovery remain the same. 
To this end, we adopt the Wayback Machine, the suggested pipeline by~\citet{jahanshahi2020Wayback}, to recreate past events.

We first extract bug fixing information from Bugzilla. 
Then, we revisit past events in the system during the training time and keep track of the information related to active developers and feasible bugs. 
When the training phase is completed, SVM and LDA learn the suitability and cost values accordingly. 
Afterwards, we continue recording the incoming bugs and their possible dependencies during the testing phase. 
Once in a day, we query the currently feasible bugs' information in the system and apply our IP model to those. 
The outcome of the model is a list of developers and assigned bugs. 
A few bugs may remain unassigned based on the imposed constraints. 
\textcolor{black}{For instance, they might be blocked by a bug that is not yet fixed, or the schedule of the suitable developer is currently full.}
We call back those bugs in the upcoming days until they meet the criteria. 
The steps of the full procedure is provided below.

\begin{enumerate}[start=0,label={(\bfseries Step \arabic*):}, leftmargin = 4em]
    \item \textbf{Setting hyper-parameters}: We use the third quartile \textcolor{black}{of bug fixing times} as the value for upper limit $\mathcal{L}$. All \textcolor{black}{$T_{jt}^d$ values have the length of} $\mathcal{L}$ and cannot exceed this limit during the process. Furthermore, SVM and LDA hyperparameters are defined similar to the previous studies. 
    
    \smallskip
    \item \textbf{Constructing SVM and LDA}: To estimate the suitability $s_i^d$ and the fixing time $c_i^d$ for developer $d$ and bug $i$, we train SVM and LDA at the end of the training phase.
    
    \smallskip
    \item \textcolor{black}{\textbf{Estimating the number of schedule slots}: Each developer may work on $j$ bugs simultaneously. The number of slots (i.e., capacity) of developer $d$ is estimated based on the developer's history of fixing bugs during the training phase. We use the third quartile and interquartile range of previous simultaneous tasks of the developers to determine the value, i.e., $j \leq \lfloor\text{Q}3 + (1.5\times\text{IQR})\rfloor$.}
    
    \smallskip
    \item \textbf{Predicting suitability and costs}: At the end of each day, we predict the suitability and cost values for all feasible, unassigned bugs. 
    
    \smallskip
    \item \textbf{Calling S-DABT}: By solving the IP model, we determine the assigned bugs and their associated developers. 
    Based on the constraints, the model postpones certain bugs while prioritizing others.
    
    \smallskip
    \item \textbf{Updating \textcolor{black}{$T_{jt}^d$}}: At the end of each day, we increment \textcolor{black}{$T_{jt}^d$} by 1 for developer $d$ while ensuring it does not exceed $\mathcal{L}$. 
    At the same time, we decrease \textcolor{black}{$T_{jt}^d$} of slot $j$ of developer $d$ by sum of their estimated $c_i^d$ values for all assigned bugs \textcolor{black}{to the slot $j$ of the developer.}
    
    \smallskip
    \item \textbf{Continuing to the next day (to Step 3)}: Once the assignment at day $t$ finishes, we move to the next day and repeat the process from Step 3. \textcolor{black}{We also keep track of the BDG statistics as well as all assigned and unassigned bugs of the day.}
    These steps continue until the end of the testing phase. 
\end{enumerate}

We assume that our method and baselines are implemented once a day. 
However, the process is generalizable, and the granularity of the time steps can be modified as needed (e.g., from daily to weekly).

\subsection{Dataset}\label{sec:dataset}
We consider four large OSS projects in our analysis. 
They are well-established projects with a sufficient number of bug reports. 
These projects were also adopted in the previous studies~\citep{Bhattacharya2010, kashiwa2020, lee2017applying, mani2019deeptriage}, making them suitable for comparative analysis. 
We collect the raw data from the repository using the Bugzilla REST API\footnote{\url{https://wiki.mozilla.org/Bugzilla:REST_API}}. 
It includes general attributes of the bugs and their metadata change history. 
Table~\ref{tab:dataset} shows the summary information for the extracted datasets. 
We use bug reports for the past two years as the testing set and the older ones as the training set. 

\begin{table*}[!ht]
\caption{\textcolor{black}{Summary information for the extracted datasets. The training phase starts from Jan. 1st, 2010 to Dec. 31st, 2017, whereas the testing phase covers Jan. 1st, 2018 to Dec. 31st, 2020.}\label{tab:dataset}}%
\resizebox{\linewidth}{!}{
    \begin{tabular}{lrr|rr|rr|rr}
        \toprule
        & \multicolumn{2}{c|}{{\scshape{{EclipseJDT}}}} &
        \multicolumn{2}{c|}{{\scshape{{LibreOffice}}}} & \multicolumn{2}{c|}{{\scshape{{GCC}}}} & \multicolumn{2}{c}{{\scshape{{Mozilla}}}} \\ 
         & \textbf{Training} & \textbf{Testing} & \textbf{Training} & \textbf{Testing} & \textbf{Training} & \textbf{Testing} &  \textbf{Training} & \textbf{Testing}  \\
        \midrule
        \textbf{Total bugs reported} & 12,598 & 3,518 & 55,554 & 14,582 & 34,635 & 9,998 & 90,178 & 22,353 \\
        \textbf{Total bug dependencies found} & 2,169 & 970 & 28,472 & 21,894 & 4,462 & 3,268 & 71,549 & 19,223\\
        \textbf{Total relevant changes in the bugs' history} & 55,109 & 15,505 & 267,310 & 94,682 & 138,580 & 42,117 & 410,010 & 114,778 \\
        \textbf{Mean and Median fixing time (days)} & (41.2, 3) & (15.7, 1) & (18.3, 1) & (9.4, 2) & (42.3, 3) & (42.0, 3) & (27.2, 5) & (12.6, 4) \\
        \textbf{Minimum, Maximum fixing time (days)} & (1, 1,753) & (1, 423) & (1, 1,484) & (1, 428) & (1, 2,396) & (1, 681) & (1, 2,172) & (1, 550) \\ 
        \midrule
        \textbf{After cleaning} &  &  &  &  &  &  &  & \\
            \qquad \textbf{0. Bugs that are not META bugs} & 12,598 & 3,518 & 54,907 & 14,416 & 34,634 & 9,996 & 89,354 & 21,912 \\
            \qquad \textbf{1. Bugs with resolved status} & 11,296 & 2,619 & 46,905 & 11,114 & 29,057 & 7,195 & 79,108 & 18,720 \\
            \qquad \textbf{2. Bugs assigned to active developers} & 3,795 & 1,491 & 5,035 & 1,888 & 27,088 & 6,719 & 10,948 & 6,365 \\
            \qquad \textbf{3. Bugs with known assignment date} & 3,021 & 1,348 & 4,345 & 1,749 & 8,628 & 3,072 & 6,768 & 3,947 \\
            \qquad \textbf{4. Bugs with acceptable fixing time} & 2,372 & 1,201 & 3,462 & 1,428 & 7,166 & 2,459 & 5,618 & 3,547 \\
        \bottomrule
    \end{tabular}
}
\end{table*}

Similar to previous studies~\citep{kashiwa2020,park2011costriage}, we consider only the bugs that meet the following \textcolor{black}{five} criteria.

\begin{enumerate}\setItemSep{0.5em}
    \item \textcolor{black}{\citet{jahanshahi2020Wayback} showed that META bugs mimic blocking bugs by providing links to other bugs via ``depends on/blocks'' mechanism. 
    Therefore, their main aim is to group similar bugs and do not have testcases of their own. 
    We exclude them as they are not a real bug reported by users to the system.}
    
   \item Not all bugs have fixing time information available. 
   For instance, some bugs are still open, or in some cases, the exact fixing date is not available in history. 
   Hence, we only consider the FIXED or CLOSED bugs with sufficient information.
   
    \item The assignment date is not available for some bugs, or the assignment time is after their resolution. 
    We eliminate them as invalid bugs.
    
    \item \textcolor{black}{We define \textit{active} developers as the ones with bug fix numbers greater than the IQR of bug fix numbers of all developers. 
    Only the bugs assigned to active developers of the training phase are considered for the testing phase.}

    \item Some bugs took years to get solved in the ITS. 
    We remove outlier bugs whose fixing time is greater than the threshold of Q3 + (1.5 $\times$ IQR), where Q3 is the third quartile of the fixing time, and IQR is the interquartile range, i.e., 
    \begin{equation*}
        \text{IQR} = (\text{the third quartile}) - (\text{the first quartile}).
    \end{equation*}
    
    The maximum acceptable fixing times that we found for {\scshape{EclipseJDT}}, {\scshape{LibreOffice}}, {\scshape{GCC}} and {\scshape{Mozilla}} are \textcolor{black}{21, 22, 38.5, and 6 days, respectively.}

\end{enumerate}

Table~\ref{tab:dataset} shows the effect of applying each filtering step on the number of feasible bugs in the system.
\textcolor{black}{Similar to previous studies~\citep{kashiwa2020,liu2016multi,park2016cost},} we preprocess the textual information through lemmatization, stop words, numbers and punctuation removal, and lengthy word elimination (i.e., longer than 20 characters). 
Finally, we merge titles and descriptions of the bugs and tokenize them.

\section{Numerical Study}~\label{sec:results}
We compare S-DABT and DABT with the well-known bug triage methods from the literature, namely, CBR using SVM, CosTriage, and RABT as well as the actual bug assignment, which is taken as a baseline. 
We implemented all the methods in Python and used Gurobi 9.5 to solve the IP models.

\subsection{Performance Discussion}
We define different metrics for the comparison as shown in Table~\ref{tab:comp_algorithms}. 
We observe that S-DABT and DABT have few unassigned bugs at the end of the testing phase due to multiple constraints imposed in their formulations. 
This number is negligible compared to the total bugs that are addressed. 
% \textcolor{black}{It uses the branch-and-bound algorithm with some heuristics to choose the incumbent (i.e., the best integer solution found at any point).}
One of the main differences between the algorithms is in their usage of available developers. 
Considering three larger projects ({\scshape{LibreOffice}}, \textcolor{black}{{\scshape{GCC}}} and {\scshape{Mozilla}}), we observe that actual assignment, RABT, DABT, \textcolor{black}{and S-DABT} use more developers than the other two methods. 
\textcolor{black}{The similarity between the manual assignment and IP-based solutions indicates that they all enforce some constraints on developers' available times.} 
CBR and CosTriage do not incorporate task concentration in their formulation, and they are still prone to over-specialization. 
We further explore this finding by reporting the average and standard deviation of the number of bugs assigned among developers. 
RABT, DABT, and S-DABT maintain the level of fair bug distribution among developers by utilizing more developers. 
\textcolor{black}{DABT and S-DABT achieve an almost similar distribution to that of RABT although they use a smaller number of developers during the testing phase. 
That is, they do not employ superfluous developers to achieve such a fair distribution (e.g., see {\scshape{Mozilla}}'s assigned developers and task distribution in Table~\ref{tab:comp_algorithms}).}

\begin{table*}[!ht]
\centering
\caption{\textcolor{black}{Comparison of different algorithms}\label{tab:comp_algorithms}}%
\resizebox{\linewidth}{!}{
    \begin{tabular}{cl>{\columncolor[HTML]{EFEFEF}}r rrrrr} %  {l|r|rrrr}
        \toprule
         & \textbf{Metrics} & \textbf{Actual} & \textbf{CBR} & \textbf{CosTriage} & \textbf{RABT} & \textbf{DABT} & \textbf{S-DABT} \\
        \midrule
       \multirow{14}{*}{\STAB{\rotatebox[origin=c]{90}{\scshape{{EclipseJDT}}}}} 
         & \textbf{\# of assigned bugs} & 1,231 & 1,231 & 1,231 & 1,231 & 1,224 & 1,226 \\
         & \textbf{\# of un-assigned bugs} & 0 & 0 & 0 & 0 & 7 & 5 \\
         & \textbf{\# of assigned developers} & 15 & 13 & 12 & 15 & 15 & 15 \\
         & \textbf{Task distribution among developers $(\mu \pm \sigma)$} & 80.0$\pm$92 & 94.7$\pm$149 & 102.6$\pm$152 & 82.1$\pm$98 & 81.6$\pm$94 & 81.7$\pm$\textbf{89} \\
         & \textbf{Mean fixing days per bug} & 6.4 & 5.5 & 5.2 & 4.8 & 4.6 & \textbf{4.3} \\
         & \textbf{Percentage of overdue bugs} & 27.8 & 71.3 & 68.2 & 10.6 & \textbf{9.2} & 12.3 \\
        %  & \textbf{Percentage of un-fixed bugs} & 27.8 & 71.3 & 68.2 & 10.6 & \textbf{9.8} & 12.7 \\
         & \textbf{Accuracy of assignments} & 97.8 & 96.9 & \textbf{97.0} & 94.6 & 90.5 & 91.1 \\
         & \textbf{Infeasible assignment w.r.t. bug dependency} & 4.2 & 5.6 & 5.5 & 3.9 & 0.7 & \textbf{0.6} \\
         & \textbf{Mean depth of the BDG} & 0.05 & 0.05 & 0.05 & 0.05 & 0.04 & 0.04 \\
         & \textbf{Mean degree of the BDG} & 0.05 & 0.05 & 0.05 & 0.04 & 0.04 & 0.04 \\
         & \textbf{Assignment divergence $(\mu \pm \sigma)$} & 0.0$\pm$0 & 82.2$\pm$235 & 82.2$\pm$235 & 75.41$\pm$233 & 74.0$\pm$233 & 75.9$\pm$234 \\
         & \textbf{Utilization} & 0.47 & 0.35 & 0.33 & 0.37 & 0.34 & \textbf{0.38} \\
         & \textbf{Running time (hrs.)} & 0.1 & 0.1 & 0.1 & 0.1 & 0.2 & 0.2 \\
         %%%%%
         %%%%%
         \midrule
        \multirow{14}{*}{\STAB{\rotatebox[origin=c]{90}{\scshape{\textbf{LibreOffice}}}}} 
        & \textbf{\# of assigned bugs} & 1,406 & 1,406 & 1,406 & 1,406 & 1,406 & 1,406 \\
         & \textbf{\# of un-assigned bugs} & 0 & 0 & 0 & 0 & 0 & 0 \\
         & \textbf{\# of assigned developers} & 51 & 17 & 17 & 60 & 56 & 60 \\
         & \textbf{Task distribution among developers $(\mu\pm \sigma)$} & 27.5$\pm$71 & 82.7$\pm$228 & 82.7$\pm$237 & \textbf{23.4}$\pm$92 & 25.1$\pm$96 & \textbf{23.4}$\pm$101 \\
         & \textbf{Mean fixing days per bug} & 2.6 & 2.1 & 2.1 & 2.0 & \textbf{1.9} & \textbf{1.9} \\
         & \textbf{Percentage of overdue bugs} & 8.7 & 39 & 52.2 & 5.1 & \textbf{4.8} & 7.2 \\
         & \textbf{Percentage of un-fixed bugs} & 8.7 & 39 & 52.2 & 5.1 & \textbf{4.8} & 7.2 \\
         & \textbf{Accuracy of assignments} & 93.4 & 99.2 & \textbf{99.3} & 93.5 & 92.4 & 93.0 \\
         & \textbf{Infeasible assignment w.r.t. bug dependency} & 0.1 & 0.1 & \textbf{0} & 0.1 & \textbf{0} & \textbf{0} \\
         & \textbf{Mean depth of the BDG} & 0.01 & 0.01 & 0.01 & 0.01 & 0.01 & 0.01 \\
         & \textbf{Mean degree of the BDG} & 0.01 & 0.01 & 0.01 & 0.01 & 0.01 & 0.01 \\
         & \textbf{Assignment divergence $(\mu \pm \sigma)$} & 0.0$\pm$0 & 73.3$\pm$148 & 73.3$\pm$148 & 73.1$\pm$148 & 73.1$\pm$148 & 73.2$\pm$148 \\
         & \textbf{Utilization} & 0.053 & 0.030 & 0.028 & 0.031 & 0.027 & \textbf{0.032} \\
         & \textbf{Running time (hrs.)} & 1 & 1 & 1.1 & 1.1 & 1.1 & 1.1 \\ 
         %%%%%
         %%%%%
         \midrule
        \multirow{14}{*}{\STAB{\rotatebox[origin=c]{90}{\textcolor{black}{\scshape{{GCC}}}}}} 
        & \textbf{\# of assigned bugs} & 2,490 & 2,490 & 2,490 & 2,490 & 2,482 & 2,482 \\
         & \textbf{\# of un-assigned bugs} & 0 & 0 & 0 & 0 & 8 & 8 \\
         & \textbf{\# of assigned developers} & 47 & 32 & 30 & 47 & 47 & 43 \\
         & \textbf{Task distribution among developers $(\mu\pm \sigma)$} & 52.3$\pm$87 & 77.8$\pm$176 & 83.0$\pm$181 & 53.0$\pm$121 & \textbf{52.8}$\pm$123 & \textbf{57.7}$\pm$124 \\
         & \textbf{Mean fixing days per bug} & 6.3 & 5.3 & 5.0 & 5.3 & \textbf{4.4} & \textbf{4.4} \\
         & \textbf{Percentage of overdue bugs} & 28.9 & 57.0 & 54.4 & 9.5 & \textbf{7.5} & 12.5 \\
        %  & \textbf{Percentage of un-fixed bugs} & 8.7 & 39 & 52.2 & 5.1 & \textbf{4.8} & 7.2 \\
         & \textbf{Accuracy of assignments} & 98.2 & 97.1 & \textbf{96.9} & 92.4 & 87.1 & 86.2 \\
         & \textbf{Infeasible assignment w.r.t. bug dependency} & 1.1 & 0.9 & 0.8 & 0.8 & \textbf{0.3} & \textbf{0.3} \\
         & \textbf{Mean depth of the BDG} & 0.11 & 0.11 & 0.11 & 0.11 & 0.11 & 0.11 \\
         & \textbf{Mean degree of the BDG} & 0.11 & 0.11 & 0.11 & 0.11 & 0.11 & 0.11 \\
         & \textbf{Assignment divergence $(\mu \pm \sigma)$} & 0.0$\pm$0 & 35.2$\pm$157 & 35.2$\pm$157 & 35.0$\pm$156 & 35.0$\pm$155 & 35.0$\pm$155 \\
         & \textbf{Utilization} & 0.369 & 0.348 & 0.331 & 0.352 & 0.278 & \textbf{0.364} \\
         & \textbf{Running time (hrs.)} & 0.6 & 0.4 & 0.6 & 0.6 & 0.6 & 0.7 \\          
         %%%%%
         %%%%%
         \midrule
        \multirow{14}{*}{\STAB{\rotatebox[origin=c]{90}{\scshape{{{\scshape Mozilla}}}}}} 
        & \textbf{\# of assigned bugs} & 3,536 & 3,536 & 3,536 & 3,536 & 3,525 & 3,527 \\
        & \textbf{\# of un-assigned bugs} & 0 & 0 & 0 & 0 & 11 & 9 \\
        & \textbf{\# of assigned developers} & 127 & 58 & 59 & 111 & 73 & 69 \\
         & \textbf{Task distribution among developers $(\mu \pm \sigma)$} & 27.7$\pm$49 & 61.0$\pm$243 & 59.9$\pm$232 & \textbf{31.9}$\pm$70 & 48.3$\pm$88 & 51.1$\pm$82 \\
        & \textbf{Mean fixing days per bug} & 6.7 & 6.3 & 6.1 & 5.8 & \textbf{3.2} & \textbf{3.2} \\
        & \textbf{Percentage of overdue bugs} & 50.6 & 73.8 & 69.5 & 14.3 & \textbf{9.5} & 12.5 \\
        % & \textbf{Percentage of un-fixed bugs} & 50.6 & 73.8 & 69.5 & 14.3 & \textbf{9.8} & 12.8 \\
        & \textbf{Accuracy of assignments} & 77.6 & 62.6 & 61.7 & 56.2 & 38.4 & 40.6 \\
        & \textbf{Infeasible assignment w.r.t. bug dependency} & 9.8 & 10.8 & 10.3 & 4.7 & 2.3 & \textbf{1.6} \\
        & \textbf{Mean depth of the BDG} & 0.09 & 0.1 & 0.1 & \textbf{0.09} & \textbf{0.09} & \textbf{0.09} \\
        & \textbf{Mean degree of the BDG} & 0.08 & 0.09 & 0.09 & \textbf{0.08} & \textbf{0.08} & \textbf{0.08} \\
        & \textbf{Assignment divergence $(\mu \pm \sigma)$} & 0.0$\pm$0 & 24.4$\pm$94 & 25.4$\pm$94 & 25.4$\pm$94 & 23.9$\pm$94 & 24.1$\pm$94 \\
        & \textbf{Utilization} & 0.16 & 0.07 & 0.07 & 0.10 & 0.07 & \textbf{0.13} \\
        & \textbf{Running time (hrs.)} & 1.5 & 1.5 & 1.5 & 1.5 & 1.5 & 1.7 \\ 
         \bottomrule
\end{tabular}
}
\end{table*}

\begin{figure}[!ht]
     \centering
     \begin{subfigure}[b]{0.48\linewidth}
         \includegraphics[width=\textwidth]{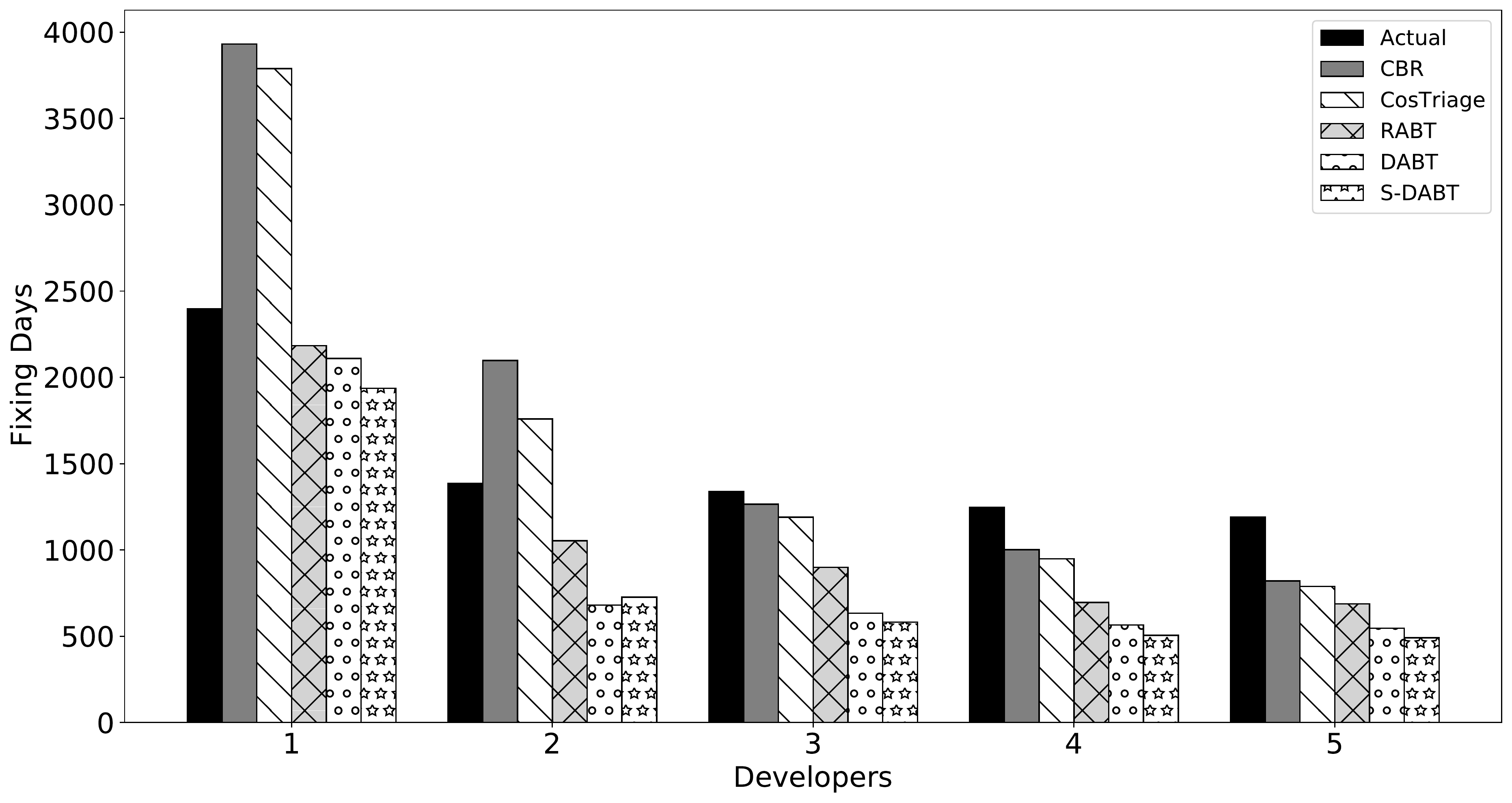}
         \caption{{\scshape{EclipseJDT}}}
         \label{fig:developer_eclipse}
     \end{subfigure}
     \begin{subfigure}[b]{0.48\linewidth}
         \includegraphics[width=\textwidth]{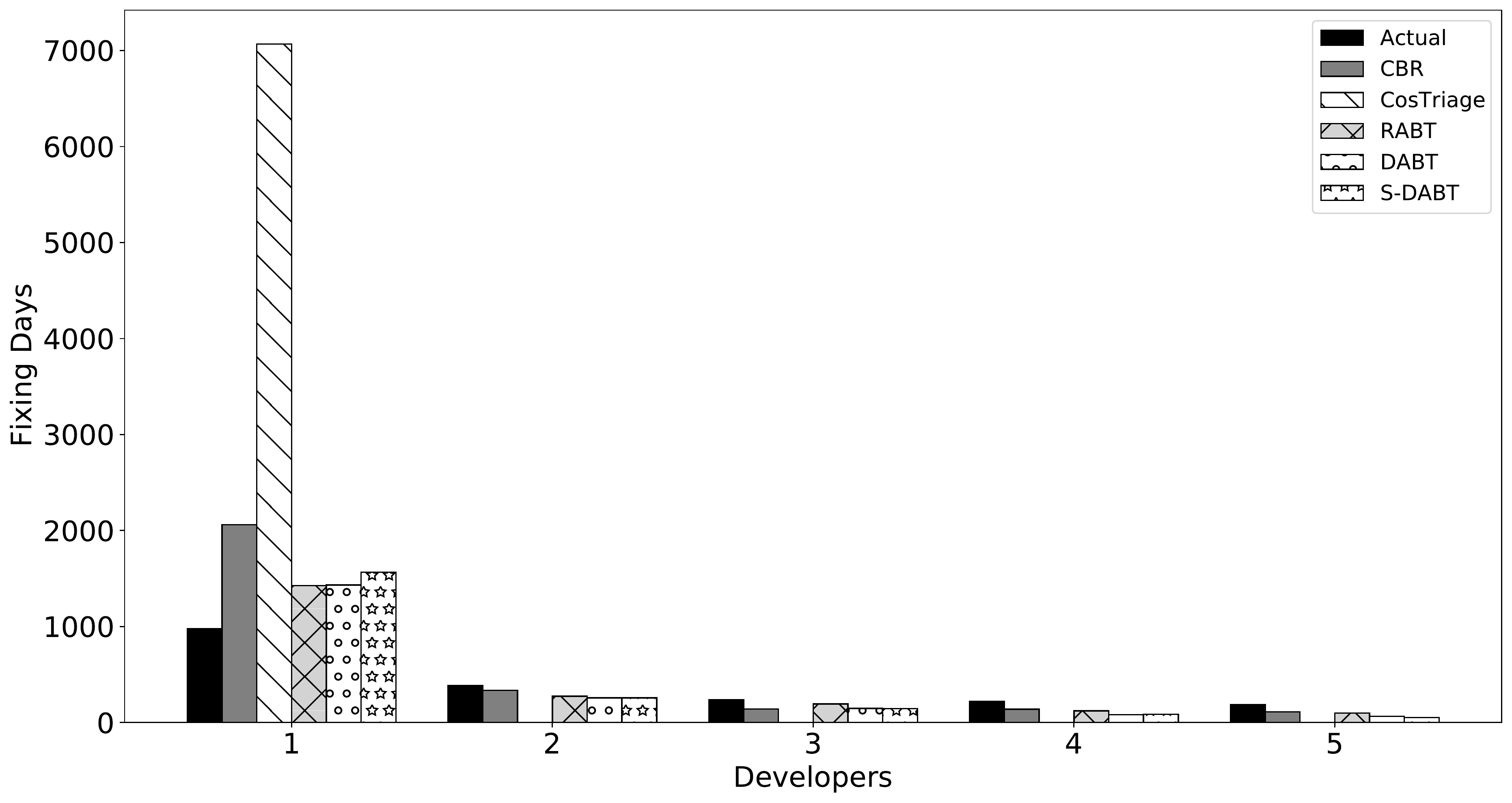}
         \caption{{\scshape{LibreOffice}}}
         \label{fig:developer_LibreOffice}
     \end{subfigure}\\
     \begin{subfigure}[b]{0.48\linewidth}
         \includegraphics[width=\textwidth]{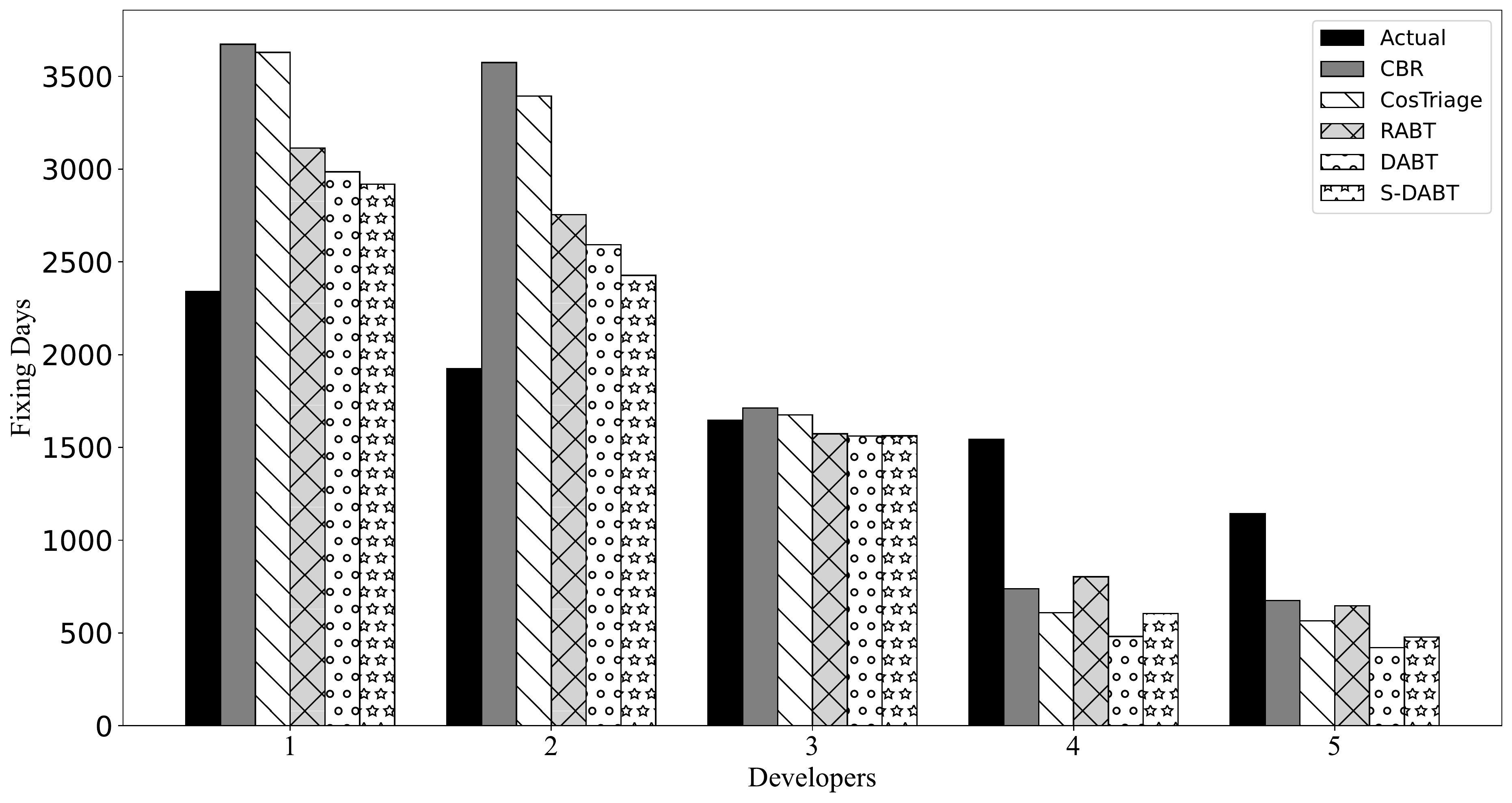}
         \caption{\textcolor{black}{{\scshape{GCC}}}}
         \label{fig:developer_GCC}
     \end{subfigure}
     \begin{subfigure}[b]{0.48\linewidth}
         \includegraphics[width=\textwidth]{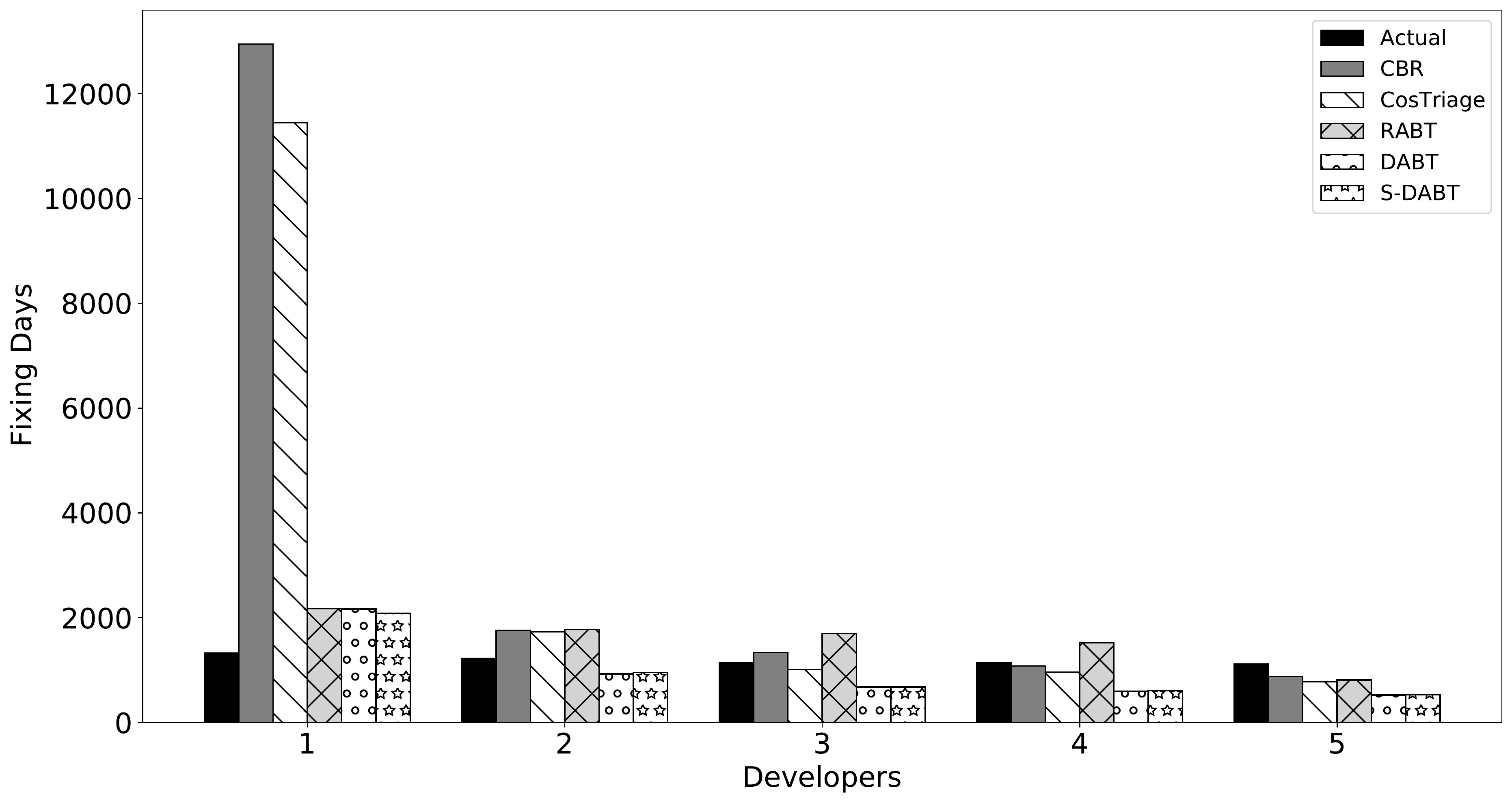}
         \caption{{\scshape{Mozilla}}}
         \label{fig:developer_Mozilla}
     \end{subfigure} 
     \caption{The number of fixing days spent by top-5 active developers during the two-year testing phase.}
        \label{fig:developer_distribution}
\end{figure}

We further investigate the effect of task concentration by examining the top-5 active developers of each algorithm.
Figure~\ref{fig:developer_distribution} shows the number of fixing days for each developer. \textcolor{black}{Note that if a developer has more than a slot to solve the bugs, the number of fixing days is the sum of the active days in all the slots.}
As CBR and CosTriage do not consider the bug fixing loads, they assign bugs to their top developer up to six times their capacity (see Figure~\ref{fig:developer_Mozilla}). 
\textcolor{black}{Although RABT, DABT, and S-DABT never over-assign bugs during the two-year testing time, S-DABT has a much lower fixing time for each developer by incorporating the fixing time minimization to the objective function and prioritizing blocked bugs.} 
Accordingly, our method both meets the schedule criterion and reduces the fixing time. 

\smallskip
\noindent\fcolorbox{black}{white}{%
    \minipage[t]{\dimexpr1\linewidth-2\fboxsep-2\fboxrule\relax}
        \textit{\textbf{RQ1-} \textcolor{black}{In general, S-DABT is able to alleviate the task concentration of the developers (i.e., we do not observe any sign of overspecialization).
        Not only does it decrease the workload of expert developers, but it also reduces the total working time of developers, with S-DABT having the lowest mean fixing time.}}
\endminipage}
\smallskip

Table~\ref{tab:comp_algorithms} shows that the average fixing days of the bugs significantly decrease when using DABT and S-DABT. 
They both consistently have the lowest fixing duration for different projects. 
\textcolor{black}{We find that S-DABT outperforms its predecessor, DABT, in {\scshape{EclipseJDT}} project.
This finding indicates the significance of using the developers' entire capacity (i.e., schedule-aware bug triage) while assigning the bugs. 
The tweak in the objective function of DABT and S-DABT compared to RABT, which enables prioritizing the bugs with shorter fixing times during bug assignments, leads to shorter bug fixing times.}
Overall, we find that DABT has the lowest rate of overdue bugs. 
When considering accumulated fixing time in the bug triage process, many bugs remain unresolved due to task concentration on particular developers.
It explains why CBR and CosTriage have a high percentage of overdue bugs, as also indicated in Figure~\ref{fig:developer_distribution}. 
By setting the hyperparameter $\mathcal{L}$, DABT does not allow the system to assign more bugs than the developers' capacity. 
On the other hand, unlike RABT, it also emphasizes the significance of the bug fixing time, leading to a lower rate of overdue bugs.
\textcolor{black}{Our proposed model, S-DABT, has a higher number of overdue bugs. 
This can be explained by the fact that S-DABT is the only model that can create detailed schedules (i.e., it assigns bugs and determines the exact fixing day). 
Therefore, when we assign bugs that should be fixed a few days later, they may exceed the release date and be counted as overdue bugs. 
However, in practice, if the ratio of delayed bugs is of importance, we recommend triagers setting a dynamic value for $\mathcal{L}$, updated by the remaining time until the next release.
Accordingly, both DABT and S-DABT will not have any unresolved bugs until the upcoming release.}

\smallskip
\noindent\fcolorbox{black}{white}{%
    \minipage[t]{\dimexpr1\linewidth-2\fboxsep-2\fboxrule\relax}
        \textit{\textbf{RQ2-} \textcolor{black}{DABT has the lowest ratio of overdue bugs by considering an upper limit on the project horizon and reducing the fixing time. On the other hand,
        S-DABT has the potential to outperform this number if the parameter $\mathcal{L}$ gets dynamic values based on the remaining time to the release. 
        Accordingly, it leads to a smoother bug fixing process and addressing more bugs before the release date.}}
\endminipage}
\smallskip

Similar to previous works~\citep{kashiwa2020,park2011costriage}, we define an accurate assignment as recommending a bug to a developer who has experience in the same component. 
Therefore, a proper bug assignment does not mean assigning to the same developer. 
According to this definition, even the manual bug assignment case fails to achieve an accuracy of 100\% when compared to the training phase. 
In practice, some developers might attempt to address a bug related to a new component for the first time. 
The component might be similar but not the same as their previous ones. 
Hence, even the actual assignment achieves an accuracy of 77.6\% for {\scshape{Mozilla}}. 
It shows that expecting a high model accuracy may cause incorrect assignment based on the ground truth values.
DABT has a lower accuracy for all projects compared to its counterparts. 
\textcolor{black}{S-DABT slightly improves the accuracy of DABT by adding the concept of developers' schedules to the assignment task and relaxing the constraint on total time at hand by adding extra working slots for some developers.}
We also note that there is a trade-off between the accuracy and speed of the bug fixing process. 
The parameter $\alpha$ in our model regulates the accuracy of the assignments and the bug fixing time. 
Moreover, we impose an upper limit on the project time horizon that may reduce the accuracy but leads to fewer overdue bugs \textcolor{black}{and prevent overspecialization}.
We further investigate the model sensitivity to parameter choices to examine whether the low accuracy values are due to the regulation parameter.

\begin{figure*}[!ht]
     \centering
     \begin{subfigure}[b]{0.40\textwidth}
         \includegraphics[width=\textwidth]{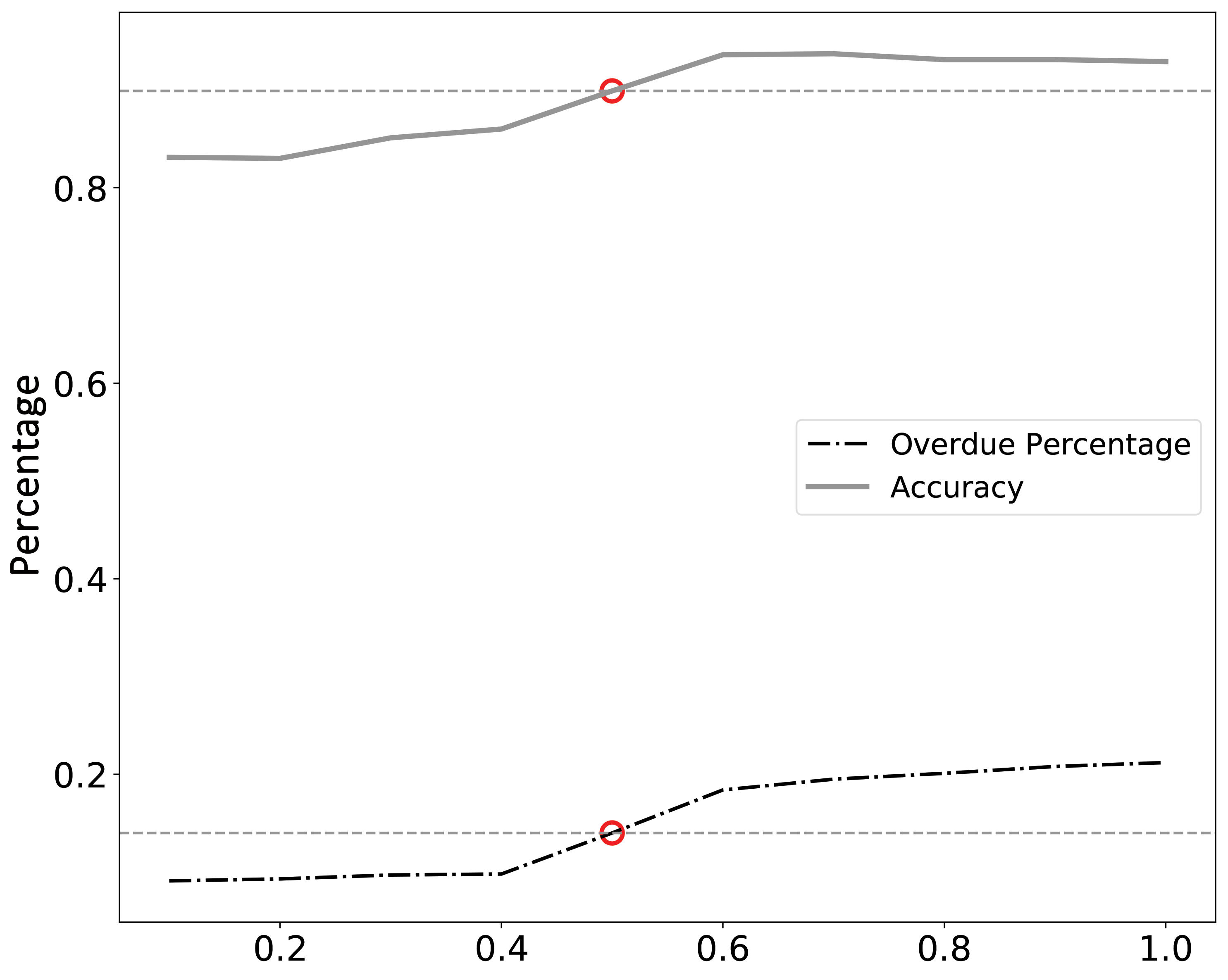}
         \caption{{\scshape{EclipseJDT}}}
         \label{fig:alpha_eclipse}
     \end{subfigure}
     \begin{subfigure}[b]{0.40\textwidth}
         \includegraphics[width=\textwidth]{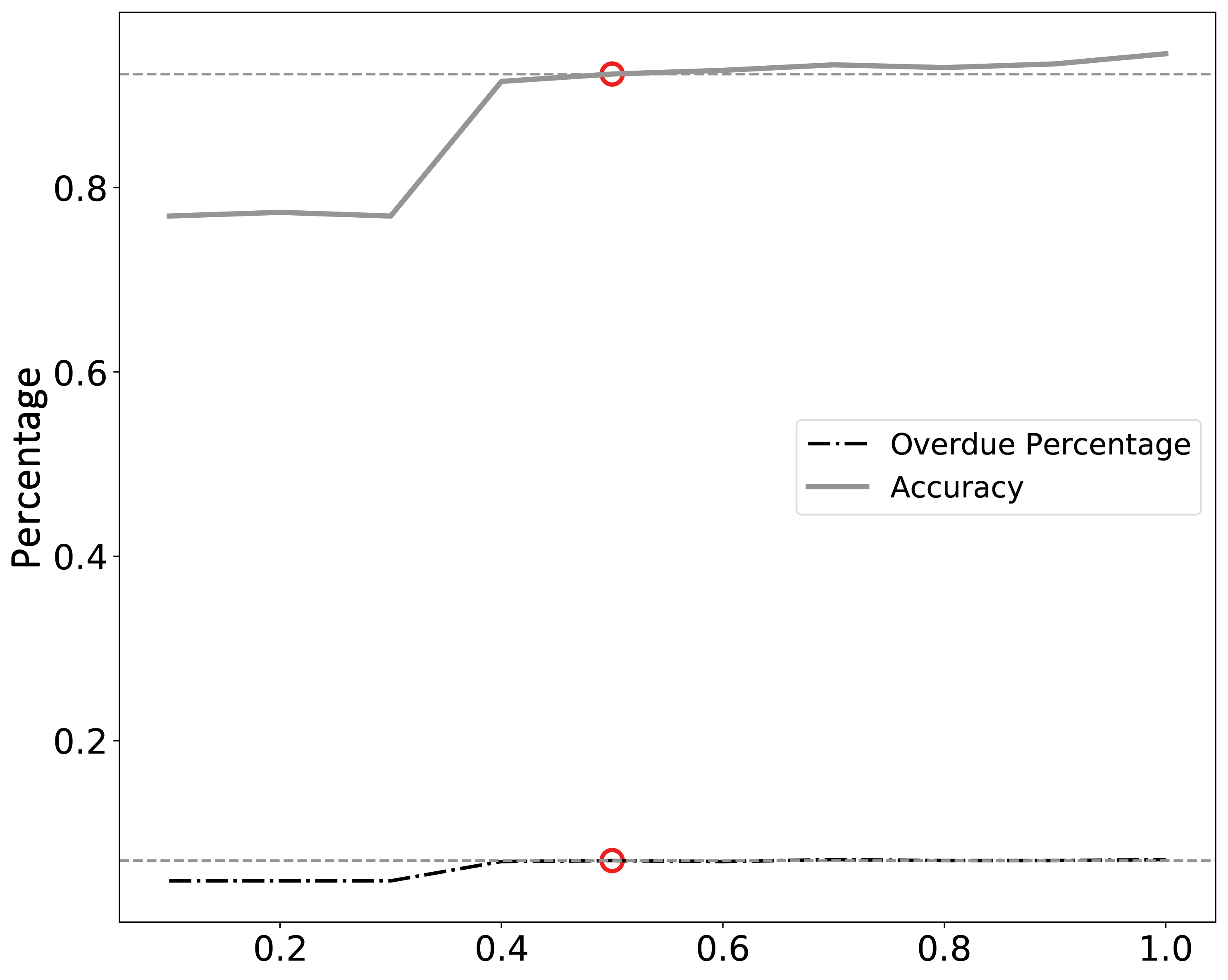}
         \caption{{\scshape{LibreOffice}}}
         \label{fig:alpha_LibreOffice}
     \end{subfigure}\\
     \begin{subfigure}[b]{0.40\textwidth}
         \includegraphics[width=\textwidth]{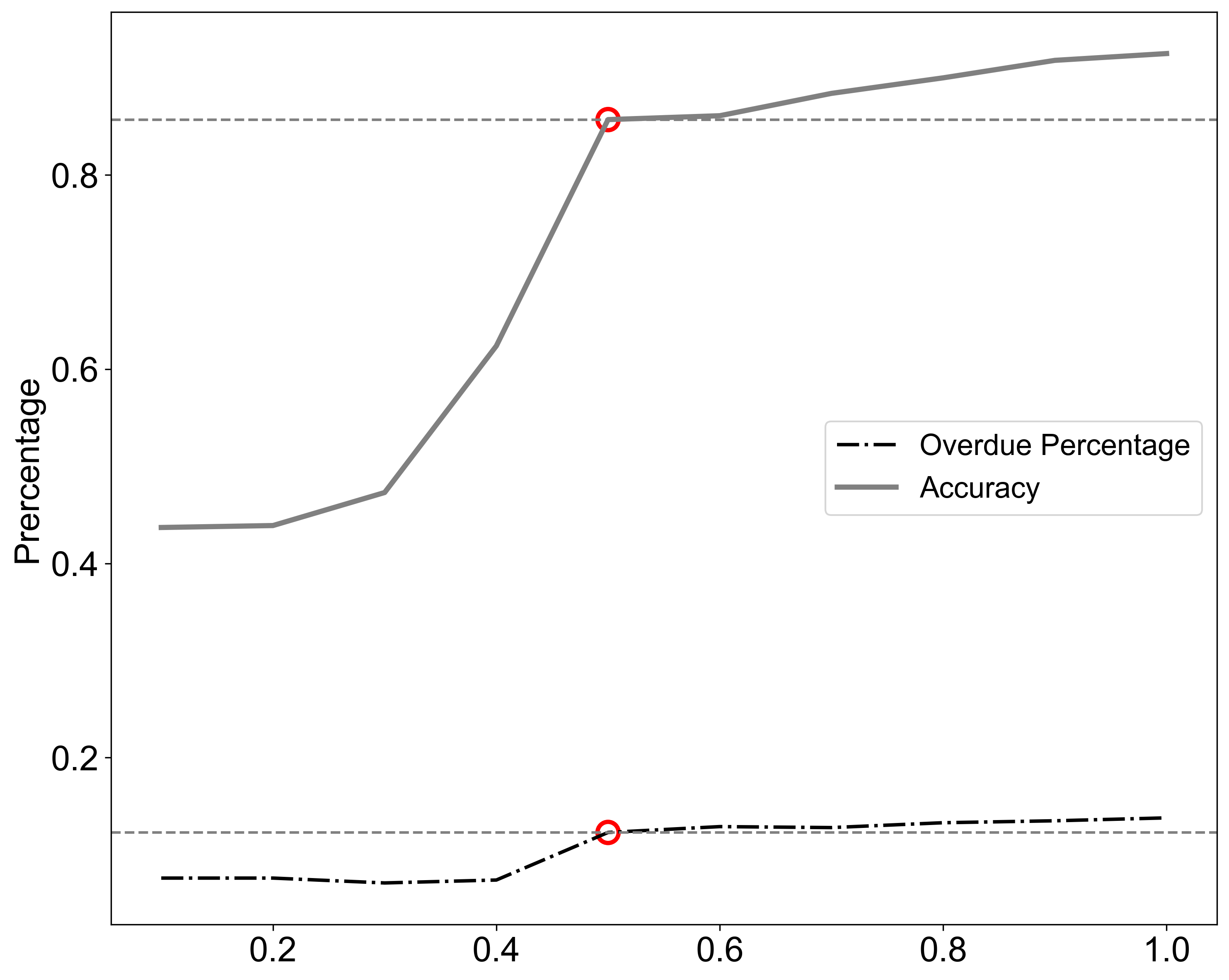}
         \caption{\textcolor{black}{\scshape{GCC}}}
         \label{fig:alpha_GCC}
     \end{subfigure} 
     \begin{subfigure}[b]{0.40\textwidth}
         \includegraphics[width=\textwidth]{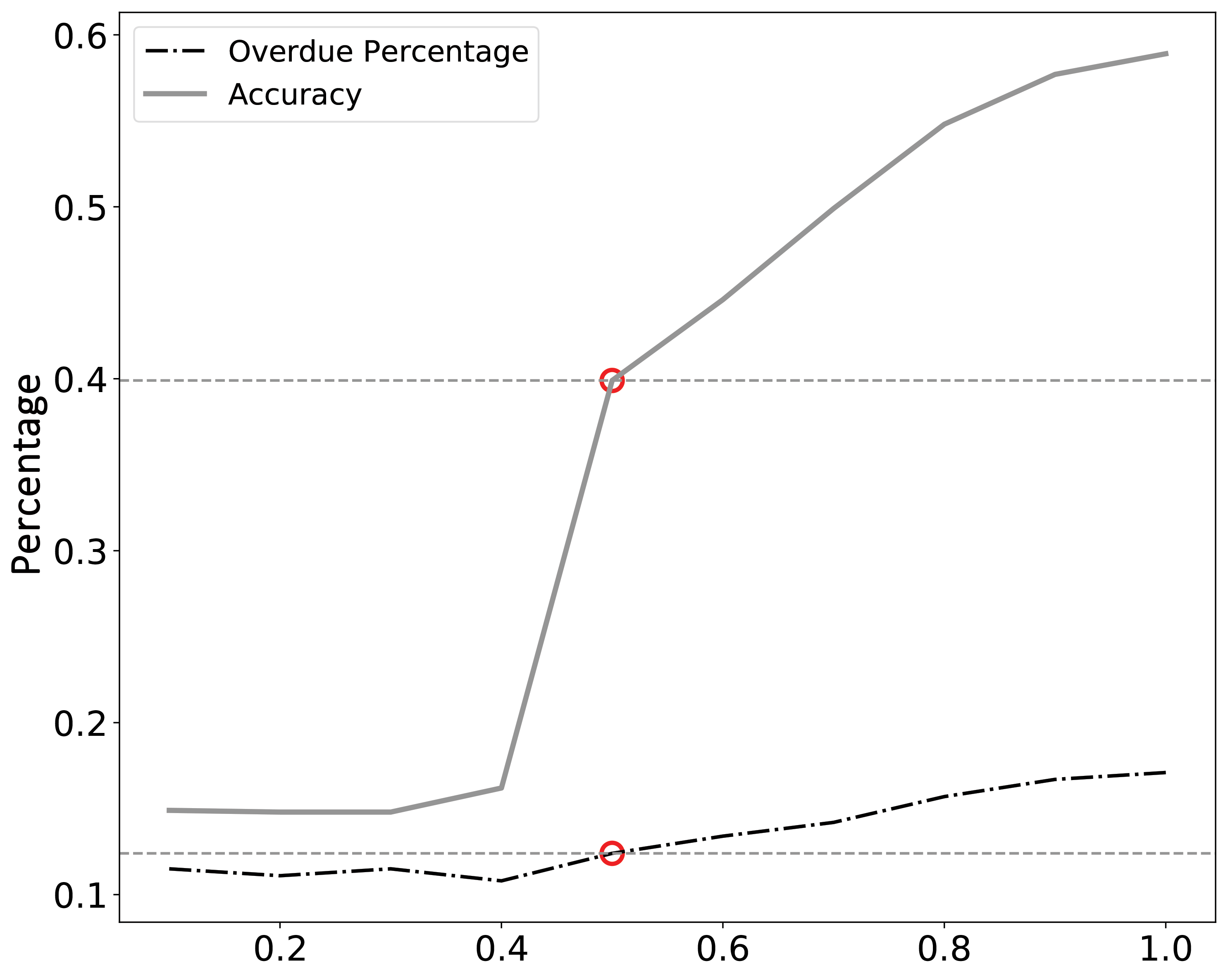}
         \caption{{\scshape{Mozilla}}}
         \label{fig:alpha_Mozilla}
     \end{subfigure} 
     \caption{Sensitivity of the accuracy and the percentage of overdue bugs of S-DABT to the $\alpha$ parameter.}
        \label{fig:sensitivity}
\end{figure*}

Figure~\ref{fig:sensitivity} shows how the accuracy and the percentage of overdue bugs are impacted by the $\alpha$ parameter.
As we increase $\alpha$, the model tends to give a higher weight to the developers' suitability than the bug fixing cost. 
Therefore, DABT achieves more accurate assignments at the expense of longer fixing times. 
In practice, the triager may decide on how much accuracy is needed, and accordingly, can set a proper value for $\alpha$.
Similar to the previous studies, we use the $\alpha = 0.5$ and give the same weight to the bug fixing time and the accuracy. 
We note that increasing the accuracy while ignoring the associated cost may result in over-specialization and task-concentration. 
\textcolor{black}{Figure~\ref{fig:alpha_Mozilla} shows that the accuracy of S-DABT can even improve by 20\% through compromising its ability to reduce the ratio of overdue bugs. We can also observe that the elbow effect of the $\alpha$ level occurs in different values. 
Specifically, $\alpha$ can be set to 0.6, 0.4, 0.5, and 0.5 for {\scshape{EclipseJDT}}, {\scshape{LibreOffice}}, {\scshape{GCC}}, and {\scshape{Mozilla}}, respectively. 
Thus, we recommend finding the ideal value for the regulation parameter based on the software project in consideration.
We further explored the possibility of improving the bug assignment accuracy. 
We examined different classification algorithms together with SVM as the baseline. 
Similar to \citet{mani2019deeptriage}'s work, we found a better prediction performance of deep learning algorithms, including LSTM and BERT}\footnote{The result can be found on our GitHub page: \href{https://github.com/HadiJahanshahi/SDABT}{https://github.com/HadiJahanshahi/SDABT}}

\textcolor{black}{One of S-DABT's key characteristics is its capability to postpone the blocked bugs and schedule bugs beforehand.} 
Whether the triager assigns a bug or a developer wants to take possession of a bug from a lengthy list of open bugs, prioritizing the blocking bugs that may increase BDG complexity is prudent. 
Therefore, S-DABT includes a constraint on the infeasible bug assignment with respect to bug dependency.
Consequently, it postpones bugs until their blocking bugs are resolved.
\textcolor{black}{Compared to DABT, S-DABT considers the dependency between the in-progress bugs (i.e., bugs already assigned but not solved) and the potential bugs to be assigned.
This additional constraint helps the algorithm to reduce the infeasible assignment due to the bug dependency.}
Table~\ref{tab:comp_algorithms} shows the better performance of S-DABT in terms of addressing blocked bugs.
\textcolor{black}{The reason why S-DABT still suffers few infeasible assignments due to the bug dependency is that there is an uncertainty in finding bug dependency~\citep{akbarinasaji2018partially}. Some dependencies might be discovered after a bug is assigned and is in the fixing process. We consider this problem as an open question to be addressed in future works.}

\begin{figure*}[!ht]
     \centering
     \begin{subfigure}[b]{0.4\textwidth}
         \includegraphics[width=\textwidth]{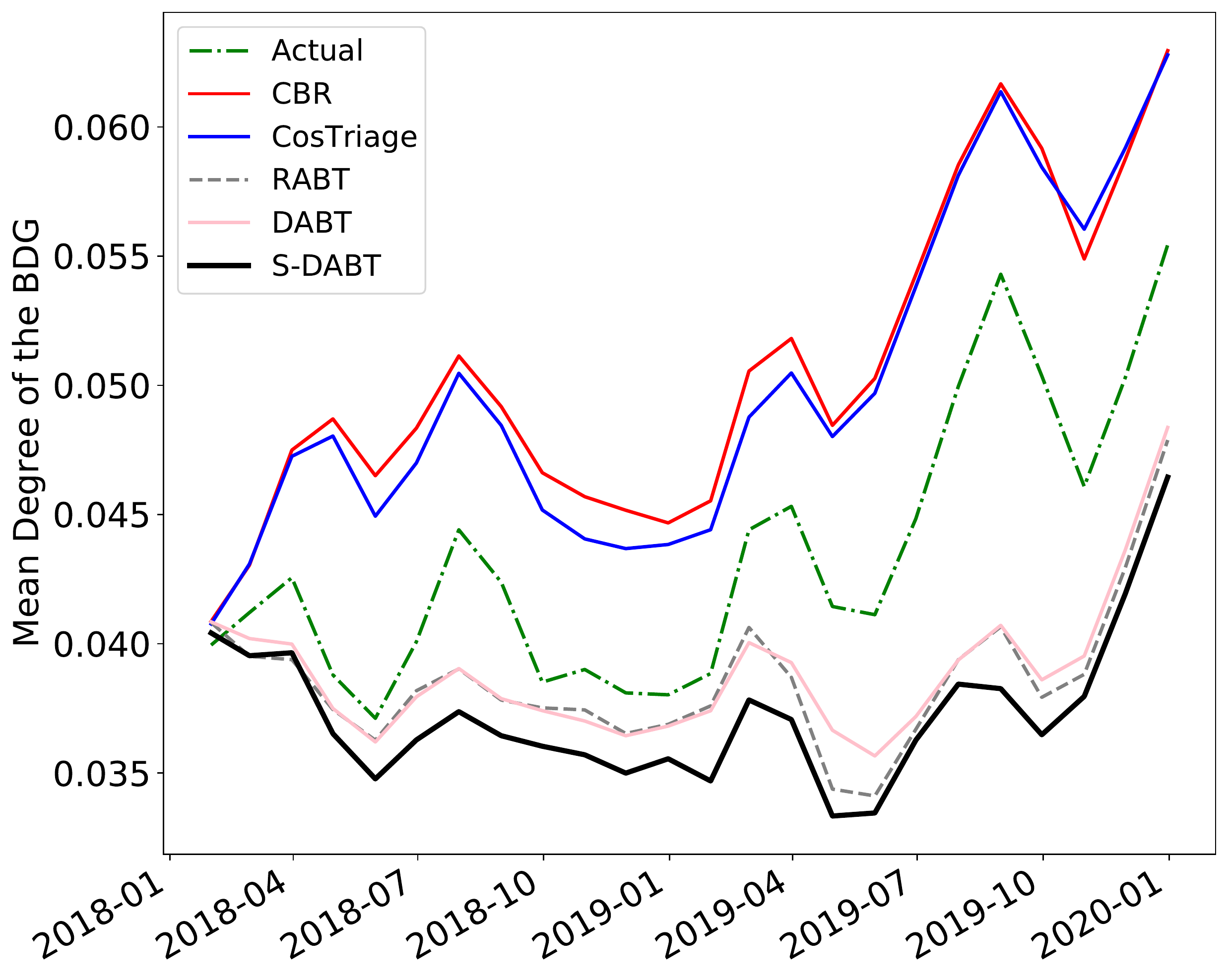}
         \caption{{\scshape{EclipseJDT}}}
         \label{fig:degree_eclipse}
     \end{subfigure}
     \begin{subfigure}[b]{0.4\textwidth}
         \includegraphics[width=\textwidth]{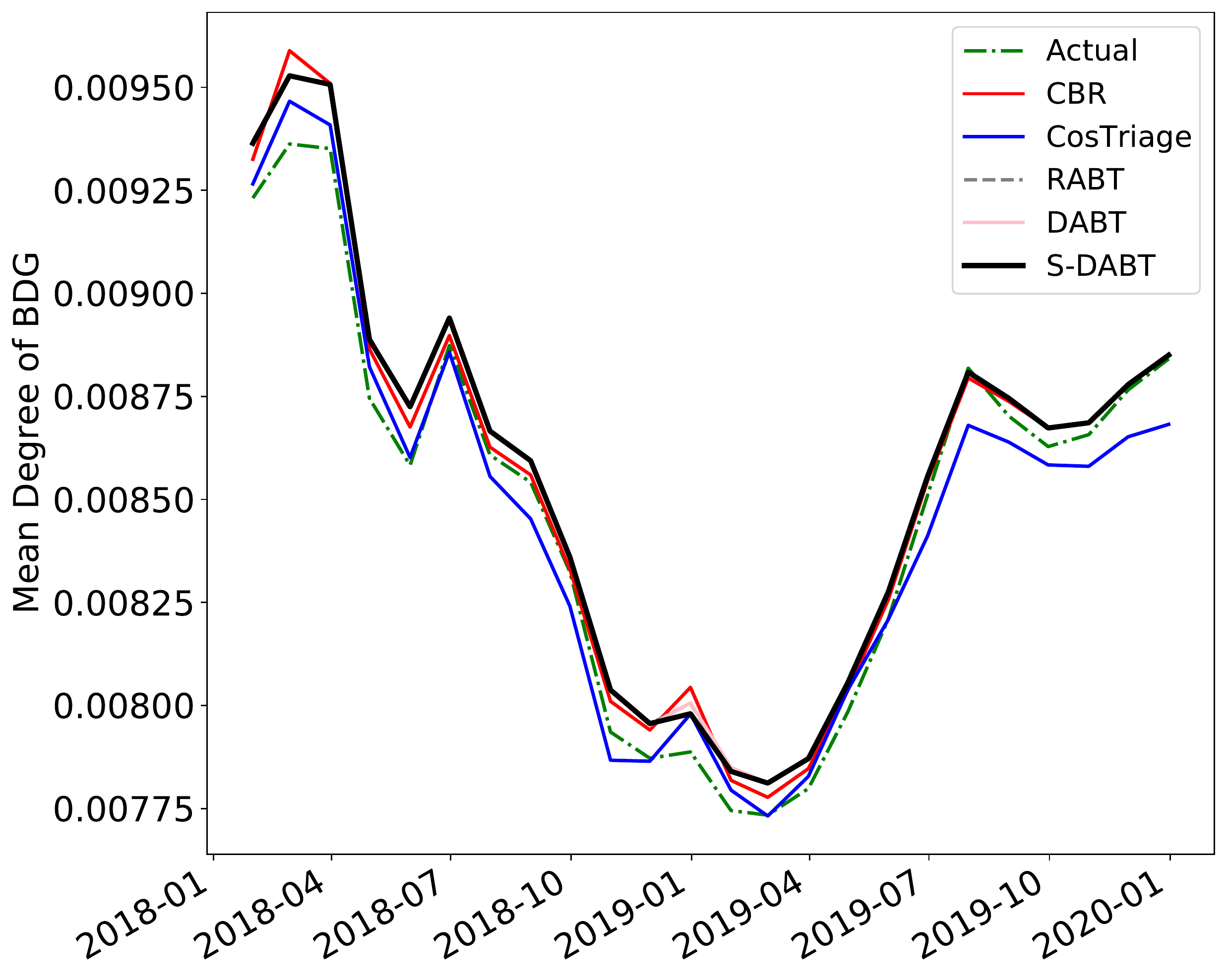}
         \caption{{\scshape{LibreOffice}}}
         \label{fig:degree_LibreOffice}
     \end{subfigure}\\     
     \begin{subfigure}[b]{0.4\textwidth}
         \includegraphics[width=\textwidth]{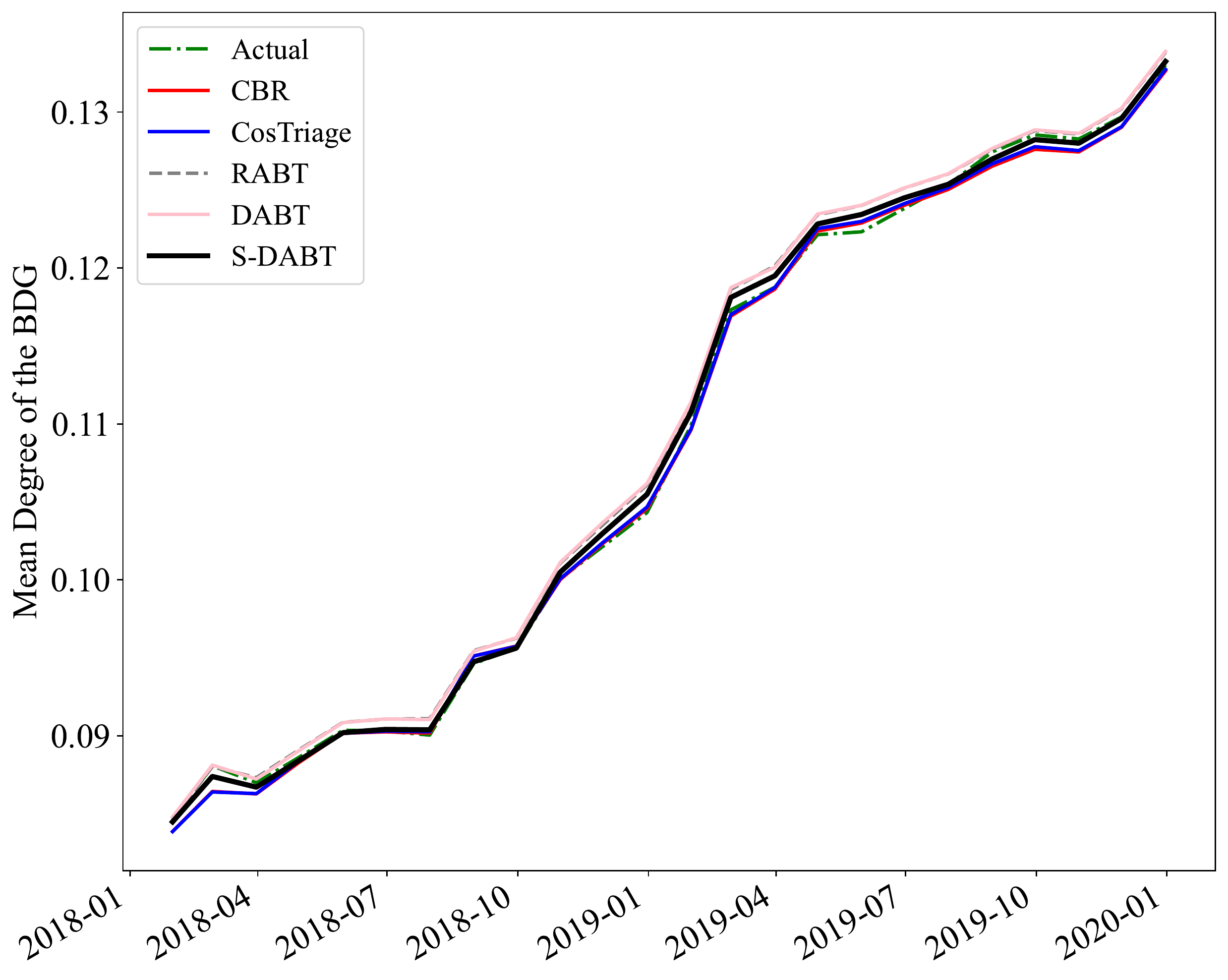}
         \caption{\textcolor{black}{\scshape{GCC}}}
         \label{fig:degree_GCC}
     \end{subfigure} 
     \begin{subfigure}[b]{0.4\textwidth}
         \includegraphics[width=\textwidth]{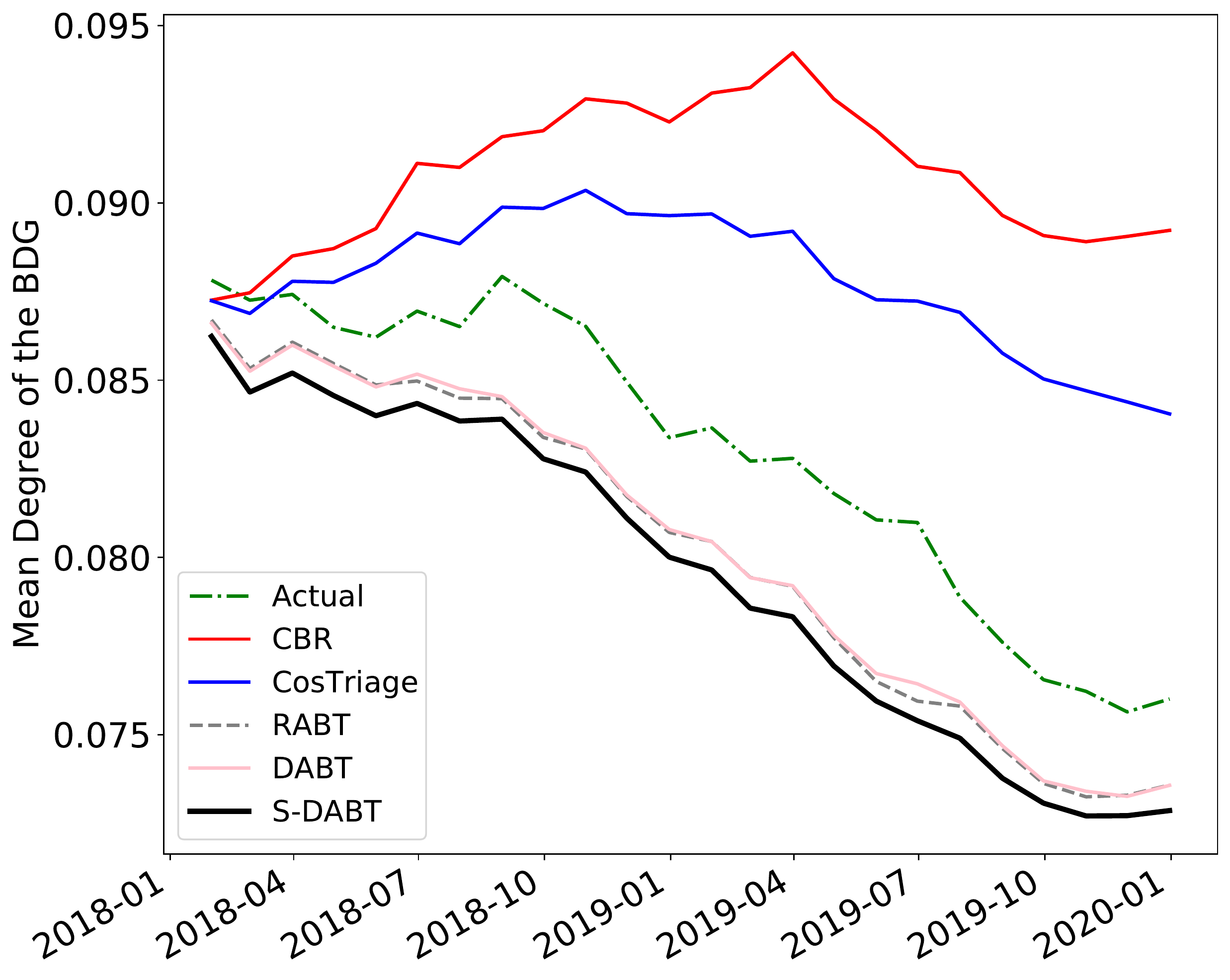}
         \caption{{\scshape{Mozilla}}}
         \label{fig:degree_Mozilla}
     \end{subfigure} 
     \caption{The effect of different strategies on the degree of the BDG during the testing phase.}
        \label{fig:degree}
\end{figure*}

We observe that DABT and S-DABT reduce the complexity of the BDG, i.e., its mean degree and depth. 
Surprisingly, RABT also shows similar performance in terms of graph complexity. 
Figure~\ref{fig:degree} explores the average degree of the bugs in the BDG during the two-year testing phase. We note that the mean degree of the BDG is already low since there are many solo bugs in the BDG. Nonetheless, a small reduction in these values can be significant since it can be considered as eliminating a few high-degree bottlenecks in the ITS.
RABT, DABT, and S-DABT keep the degree and the depth of the graph low, given all algorithms solve the same list of bugs.
For the exceptional case of {\scshape{LibreOffice}}, \textcolor{black}{our findings are consistent with that of \citep{jahanshahi2020Wayback}, which reported few bug dependencies in the project.}
Therefore, based on the other projects, we conclude that even when addressing the same list of bugs, the proper timing will reduce both the complexity of the BDG and the number of overdue bugs. 
\textcolor{black}{Particularly, we further explore the significance of the difference in graph complexity between DABT and S-DABT in Section~\ref{sec:DABT-S-DABT}.}
We note that lower BDG complexity is beneficial in the long run when the rate of incoming bugs is increasing while many bugs are still blocked by the older ones. 

\smallskip
\noindent\fcolorbox{black}{white}{%
    \minipage[t]{\dimexpr1\linewidth-2\fboxsep-2\fboxrule\relax}
        \textit{\textbf{RQ3-} \textcolor{black}{S-DABT, similar to DABT and RABT is able to reduce the complexity of the bug dependency graph} through the proper timing of bug assignment. It mitigates the risk of having a high number of blocking bugs in the long run.}
\endminipage}
\smallskip

\textcolor{black}{Table~\ref{tab:comp_algorithms} shows the assignment divergence, defined as the difference between the actual assignment time and the models' assignment time. 
It indicates a considerable gap in the literature in determining the exact assignment date for a bug.
On average, the models are early/late in assigning the bugs by approximately 78, 73, 35, and 24 days for {\scshape{EclipseJDT}}, {\scshape{LibreOffice}}, {\scshape{GCC}}, and {\scshape{Mozilla}}, respectively.
None of the models formulates or estimates the assignment date of the bugs. 
Therefore, future research may explore the feasibility of estimating the bug assignment timing.
}

\textcolor{black}{We define developer utilization as the percentage of possible developers who are active each day.
Assuming we have 100 developers, if three of them are working on the assigned bugs for the day $t$, then the developer utilization will be 3 percent.
% Accordingly, we show developers' utilization in Table~\ref{tab:comp_algorithms}. 
Utilization values in Table~\ref{tab:comp_algorithms} show that when we have fewer developers and a higher number of reported bugs, as in {\scshape{EclipseJDT}}, the utilization is higher (most of them are busy addressing the new bugs).
We note that using all the available resources in a project can be crucial.
Among all models, S-DABT is more similar to the actual case in terms of developer utilization (i.e., it has the highest utilization rate).
Considering developers' schedules in the model formulation makes it a suitable algorithm in terms of utilization.
It enhances DABT by expanding the variable definition to embrace developers' slots. 
This modification leads to an increased model complexity; however, our experimental results indicate no significant difference in the running time of DABT and S-DABT (see Table~\ref{tab:comp_algorithms}).
}
\smallskip
\noindent\fcolorbox{black}{white}{%
    \minipage[t]{\dimexpr1\linewidth-2\fboxsep-2\fboxrule\relax}
        \textit{\textbf{RQ4-} \textcolor{black}{S-DABT considers developers' schedule, making it a proper algorithm in terms of developers' utilization. It results in a fair distribution of the task in the system and the highest utilization rate compared to the previous algorithms.}}
\endminipage}
\smallskip

\subsection{\textcolor{black}{Comparison of DABT and S-DABT}} \label{sec:DABT-S-DABT}
\textcolor{black}{
    S-DABT has a more intricate structure compared to its predecessor, DABT, since it encompasses developers' schedules and a more sophisticated version of bug dependency checks. In this section, we discuss whether this additional level of complexity practically enhances the performance of the bug triage task.}

\textcolor{black}{\paragraph{Impact on bug fixing costs} 
    During the bug triage process, the capacity of each developer may impact the bug assignment decisions. If a triager assigns a bug to someone who is not currently available or has many previously unresolved bugs, this might lead to bug accumulation in the system. Accordingly, the fixing cost of a lingering bug may increase over time as it remains unaddressed in the issue tracking system. Such a cost does not only involve software development costs but also involves the opportunity cost of leaving the bug open in the system ~\citep{LingeringbugsShirin2017}. Thus, S-DABT better addresses the actual decisions in the ITS by considering the real-time schedules of the developers.}

\textcolor{black}{
    Previous studies approximated the bug fixing costs according to the time it takes to fix a bug since it is reported to the ITS. However, a developer might have been working on more than one bug while fixing this particular bug. Accordingly, the bug fixing times are affected by simultaneous bug fixing. \citet{Zaidi2022} argue that the bug triaging methods such as DABT are limited by the assumption that each developer can only work on one bug report at a time, which is not a realistic scenario in practice. As such, we enhance DABT by incorporating the capability of the developers in fixing multiple bugs simultaneously. S-DABT formulates the capacity of each developer based on their number of slots, and accordingly, its estimation of bug-fixing time becomes more realistic.}

\textcolor{black}{\paragraph{Automating the entire assignment process} 
    One of the issues with DABT is its manual reordering of the assignments after the bugs are assigned to developers. As the integer programming model in DABT does not determine the exact assignment time (e.g., day $t$) in its formulation, if the model delegates more than a bug to a developer, it does not specify which one should get fixed first. Hence, while using DABT, we are required to sort the bugs after the assignment, leading to an increase in task complexity. Decision variables $x^d_{ijt}$ in the S-DABT model directly account for such decisions and help determine the exact day on which a bug fixing should get started. Therefore, the new model automates the bug assignment in one step without a need for further manual post-processing.}

\begin{figure}[!ht]
    \centering
    \subfloat[{\scshape EclipseJDT} \label{fig:utilization_Eclipse}]{\includegraphics[width=0.4\textwidth]{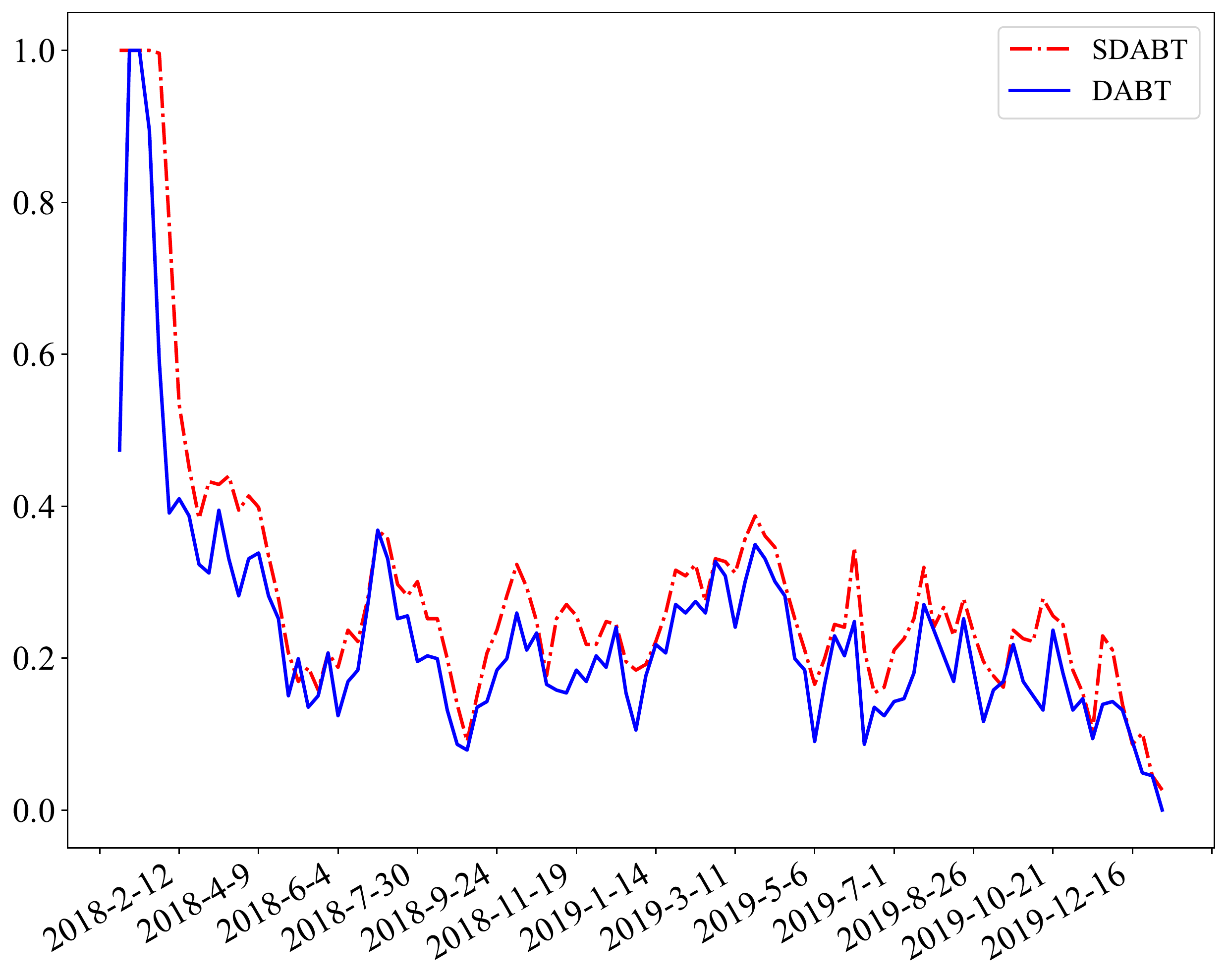}}
    \subfloat[{\scshape LibreOffice} \label{fig:utilization_LibreOffice}]{\includegraphics[width=0.4\textwidth]{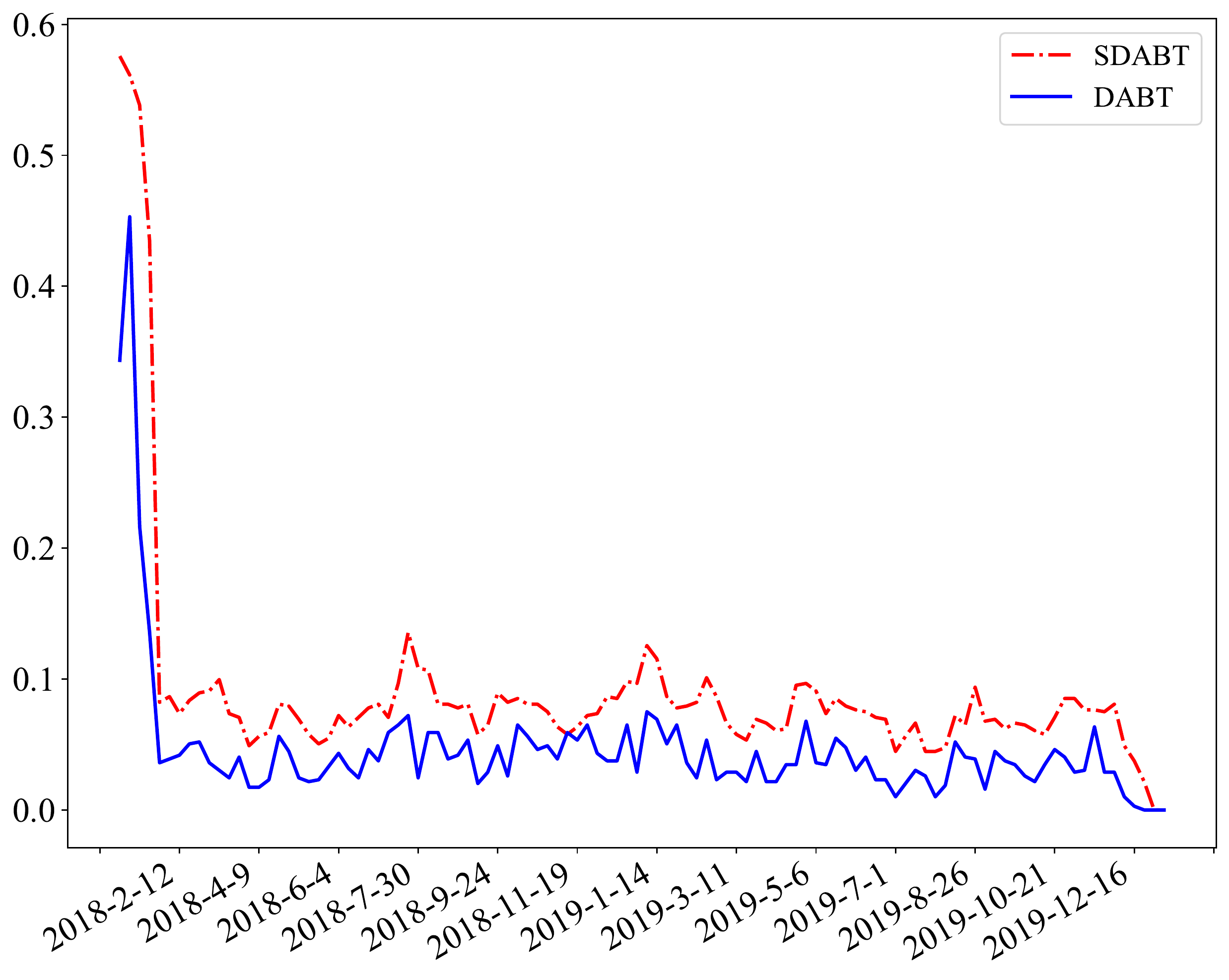}}
    \\
    \subfloat[{\scshape GCC} \label{fig:utilization_GCC}]{\includegraphics[width=0.4\textwidth]{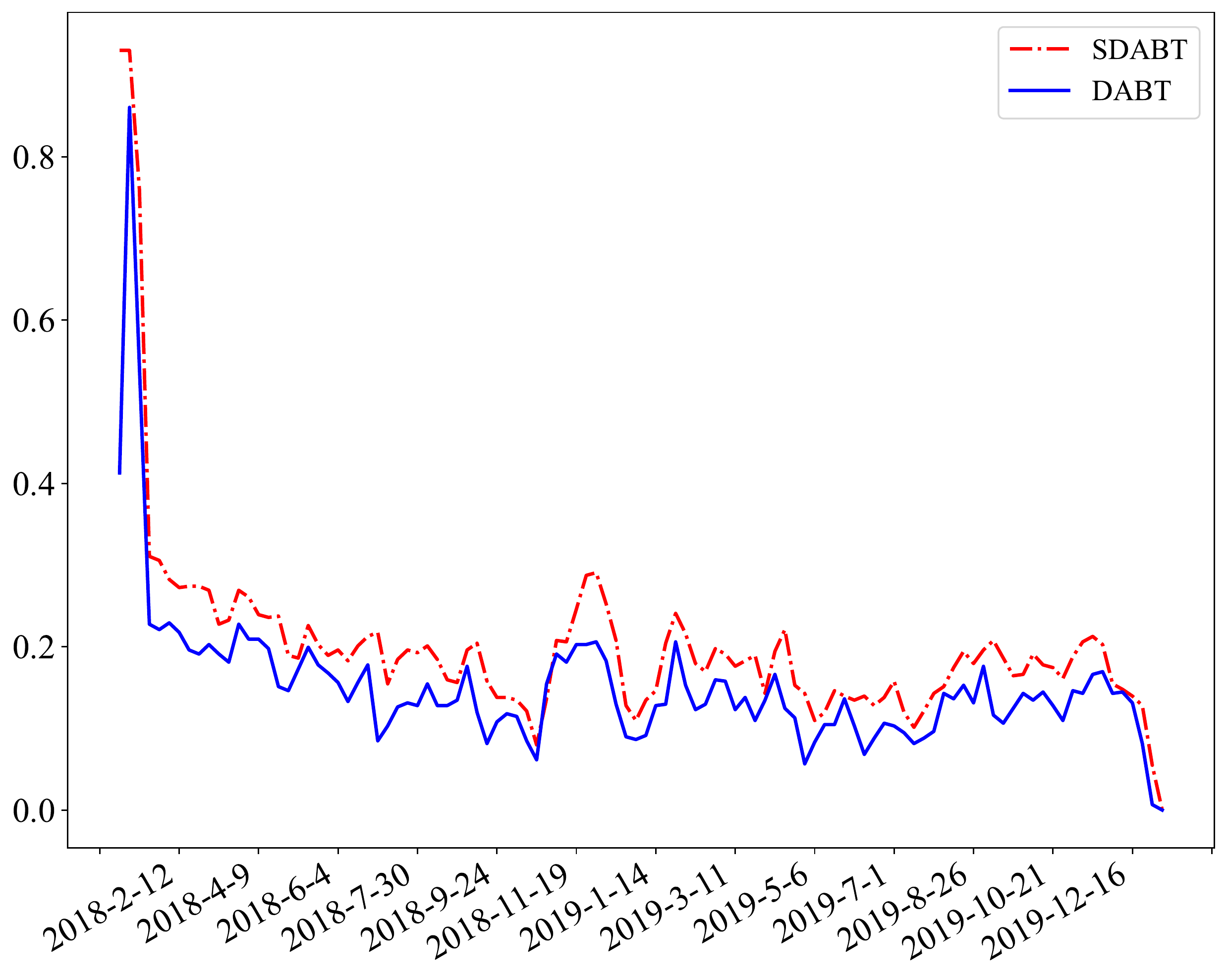}}
    \subfloat[{\scshape Mozilla} \label{fig:utilization_Mozilla}]{\includegraphics[width=0.4\textwidth]{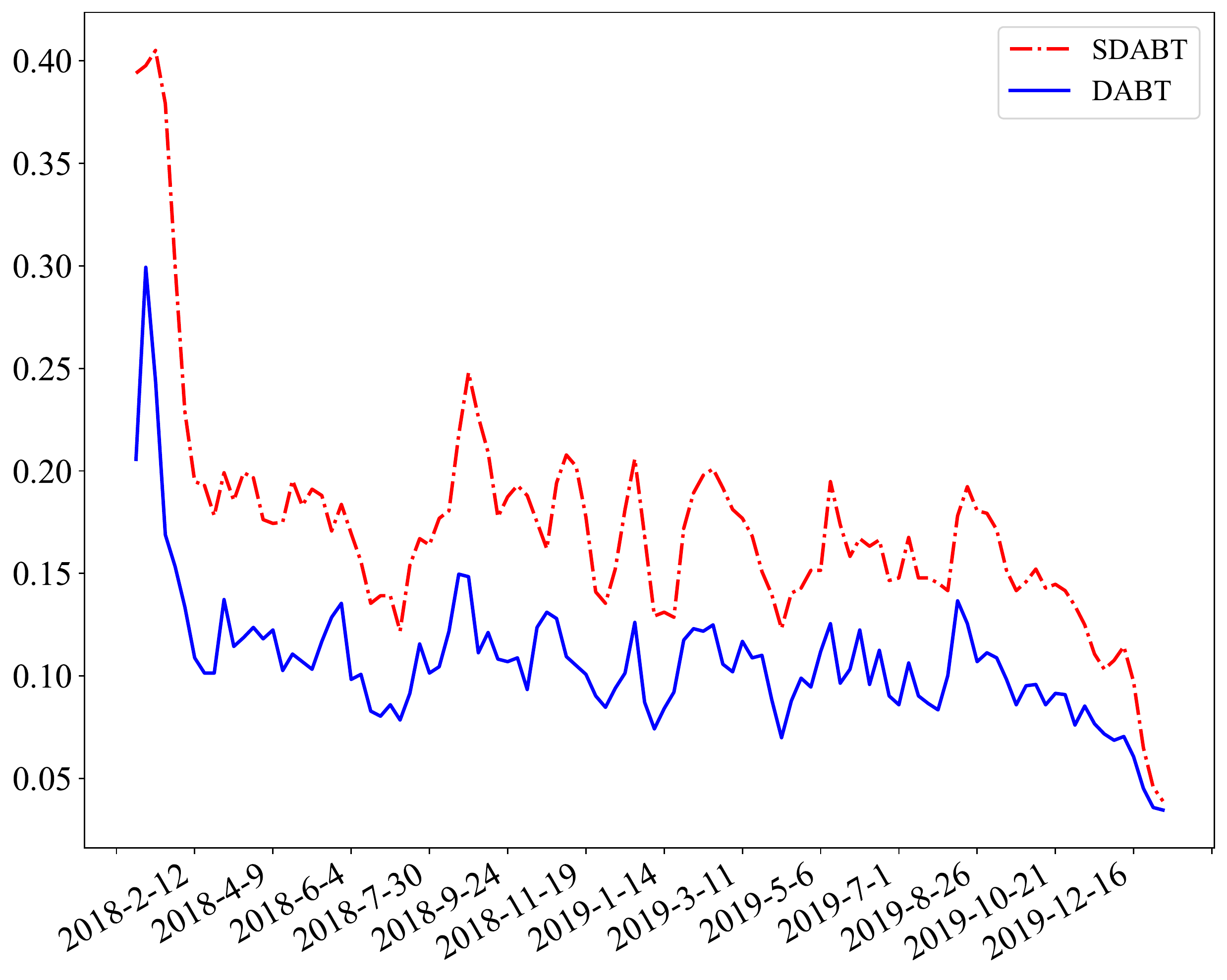}}
    \caption{Utilization values for DABT and S-DABT}
    \label{fig:utilization}
\end{figure}

\textcolor{black}{\paragraph{Efficiently utilizing developers' time} 
    We define a developer's utilization as the percentage of assigned slots out of their whole capacity. Assuming a developer can address more than a bug simultaneously, we evaluate whether our new model is capable of fully utilizing one's capacity or not. The utilization values range between 0 and 1. The higher the utilization value is, the more efficient the model is in exploiting developers' time. We track the capacity of the developers and their assigned tasks during the testing phase. Figure~\ref{fig:utilization} shows how S-DABT and DABT leverage the developers' capacity during the testing period. We average the utilization of the developers and plot them weekly. These results indicate higher utilization values for S-DABT compared to DABT. In addition, we perform a statistical test to understand whether this difference is significant. As the p-values of the Shapiro-Wilk test show, the data do not follow a normal distribution. Therefore, we utilize the non-parametric paired Wilcoxon signed-rank test with the alternative hypothesis that S-DABT has a greater utilization than DABT. Small p-values for the four projects indicate a significant utilization enhancement for S-DABT over DABT ($p=$4.8e-18, $p=$4.4e-19, $p=$3.9e-19, and $p=$1.9-19 for {\scshape Eclipse},  {\scshape LibreOffice}, {\scshape GCC}, and {\scshape Mozilla}, respectively).  }

\textcolor{black}{\paragraph{Planning ahead} 
    One of the main characteristics of the S-DABT is its ability to assign bugs ahead of time. As the model distinctively formulates both the start time of bug fixing and developers' capacities, it can assign a bug based on the future availability of a developer. In such a case, developers know which bug should be fixed each day, and also, the bug reporters (i.e., users) are aware of the assignees even before the bug fixing period gets started. Such transparency in the system enhances its reliability from both user and developer standpoints. Unlike previous models, S-DABT is designed to handle such situations; nevertheless, we test if it delivers the same experience in practice. We track the number of assigned bugs on a daily basis in the system and observe whether these values are higher for S-DABT compared to DABT. We use the non-parametric paired Wilcoxon signed-rank test to assess the statistical significance of differences between DABT and S-DABT. During the testing phase, we find a significantly greater number of assigned bugs by S-DABT than DABT (with the p-values of 1.1e-37, 1.3e-68, 0.0003, and 5.1e-111 for {\scshape Eclipse}, {\scshape LibreOffice}, {\scshape GCC}, and {\scshape Mozilla}, respectively). Hence, the new formulation reduces the number of unassigned bugs in the system, indicating that the model plans ahead in bug assignment and provides a longer-term planning perspective.}

\textcolor{black}{\paragraph{Lowering the BDG complexity} 
    S-DABT imposes an intricate set of constraints on bug dependencies, including both assigned and open bugs. Therefore, it models the complex bug dependency network more comprehensively. Using the non-parametric paired Wilcoxon signed-rank test, we examine whether the degree of the BDG decreases when S-DABT is used instead of DABT. We find that, as also shown in Figure~\ref{fig:degree}, there is a significant reduction in the number of dependencies in all projects (with the p-values of 8.3e-11, 1.19e-07, 0.03, and 0.0006 for {\scshape Eclipse}, {\scshape GCC}, and {\scshape Mozilla}, respectively). Thus, we conclude that S-DABT is able to lower the BDG complexity given that the number of reported dependencies is large enough. That is, in the {\scshape LibreOffice} project, where the number of reported dependencies is lower~\citep{jahanshahi2020Wayback}, we do not find any significant difference between the two models.}

\textcolor{black}{\paragraph{Impacts on other performance metrics} 
    We do not find any significant improvement in other performance metrics for S-DABT compared to DABT. Both models have the same objective function; therefore, it may not be possible to achieve a significant enhancement in terms of the suitability of the assigned bugs or the bug fixing costs. We note that they both indirectly lower the task concentration, and DABT and S-DABT do not dominate each other across the projects. DABT reduces the overdue bugs compared to S-DABT as it does not plan ahead. However, if we dynamically set the S-DABT's project horizon to match the release dates, we expect to achieve similar results in this aspect. As a result, we do not observe a significant difference between these two models based on these standard metrics. }

\smallskip
\noindent\fcolorbox{black}{white}{%
    \minipage[t]{\dimexpr1\linewidth-2\fboxsep-2\fboxrule\relax}
        \textit{\textcolor{black}{\textbf{RQ5-} S-DABT is able to plan ahead by determining the exact fixing day of each assigned bug. It does not require manual adjustment after the assignment as needed in DABT; thus, it fully automates the bug assignment process. Our numerical results indicate a better utilization of the developers by S-DABT and a reduction in the BDG complexity compared to DABT.}}
\endminipage}
\smallskip

\section{Threats to Validity}~\label{sec:threats}
Threats to the validity of our empirical study are as follows.

\subsection{Construct Validity}
We estimate the performance of the models using a train-test split. 
The first eight years are adopted as the training set and the last two years as the test. 
However, the evolving nature of the bug repository may have an impact on the results. 
In some cases, developers become inactive after some time or become more focused on a specific project component.
Therefore, the definition of the \textit{active} developer might be required to get updated from time to time. 
However, to make our results comparable with those of previous studies, we choose to rely on its common practice and definition. 
We recommend a rolling approach for the train-test split to overcome outdated decisions for future research. 

Although we study textual information as our independent feature, other bug characteristics --e.g., the number of comments, keywords, and the number of CC'ed developers-- can be added to the prediction models. 
We plan to expand our independent variables and include additional external factors in future works.

\textcolor{black}{To the best of our knowledge, this work is the first study to consider developers' schedules (i.e., simultaneous working slots) in the modeling. Accordingly, comparing the S-DABT's performance with that of previous algorithms is a challenging task. For IP models (e.g., RABT and DABT), we assume the capacity of a developer is equal to the maximum available time in all the developer's slots. For other models, we distribute the assigned bugs to the slot with the maximum available free days. Accordingly, our model and previous ones become comparable since all can use all developers' slots.} \textcolor{black}{Regarding the accuracy of the capacity estimation based on the bugs' history, we acknowledge the lack of data availability related to the developers' actual schedule. The same threat applies to bug fixing time estimation as well. The ground truth for bug fixing time does not include the real effort a developer invests in its resolution. A developer may work on multiple bugs at the same time, but it may increase the fixing time of all the bugs for that developer. Accordingly, since our model simultaneously utilizes bug fixing cost, suitability, and capacity, the issue of a developer having a higher capacity (i.e., solving multiple bugs at a time) can be mitigated due to potentially having a higher fixing time. As the model does not solely rely on schedule availability, the benefits of assigning multiple simultaneous bugs to a single developer are weighed against suitability and cost objectives.}

\subsection{Internal Validity}
The bug information is extracted from the Bugzilla using the REST API\footnote{\url{https://wiki.mozilla.org/Bugzilla:REST_API}}. 
We incorporate all the bug records between January 2010 and December 2019 to have updated bug report information. 
However, the API is limited for ordinary users, and access to the information of several bugs is not feasible.
Accordingly, we extract all bug comments at hand and complete our dataset using regular expressions. 
We ensure that our dataset incorporates all publicly available bugs for all four projects.

\textcolor{black}{
    We do not consider the severity and priority of the bugs in our model formulation as they are typically considered to be subjective~\citep{akbarinasaji2018partially, Gupta2019, tian2015, Yang2017}. Furthermore, we note that solely focusing on priority/severity would lead to assigning a high-impact bug to a non-expert developer from a particular category. However, our model indirectly accounts for priority and severity using two parameters in the objective function: cost and suitability. We expect to observe high cost and low suitability values for a case in which a developer is not an expert in a particular domain. Therefore, the model may postpone these bugs to future decision epochs until a more suitable developer becomes available. Another salient point to note is the relation between severity, priority, and fixing time. Previous studies pointed to faster average fixing times for those bugs with higher severity and priority~\citep{sepahvand2020, Zhang2013}. Therefore, priority and severity considerations are in a way taken into account in our models using the bug fixing times (e.g., with high priority/severity leading to shorter fixing times as observed in the OSS). 
}

\textcolor{black}{
    Previous studies indicated that blocking bugs may take longer to get fixed compared to non-blocking ones~\citep{Valdivia2018}. 
    This might imply that any automated bug triage method such as DABT and S-DABT would prioritize fixing solo bugs over bugs with dependencies.
    However, our model has strict constraints for blocking bugs that help prioritize them over the blocked bugs or solo bugs. 
    In addition, the fixing cost is only one of the parameters in the model that contributes to prioritizing bugs with a shorter fixing time; that is, the assignment variables should satisfy all the constraints to assign a bug to a particular developer. 
    Note that even if a blocking bug has a long fixing time, the suitability objective would still help prioritize such a bug so that the blocked bugs would be eligible to be resolved in future time steps. In addition, shorter fixing times for the blocked bugs might push for solving the blocking bugs first. In summary, our model has several mechanisms to prevent a bug backlog for the bugs with dependencies. This is also in line with our observations through numerical study, and we do not observe blocking bugs lingering in the system for a long time in any studied projects.
}

\subsection{External Validity} 
We consider four well-established projects in Bugzilla.
Although it is compatible with the previous works, our result may not be generalizable to all other open software systems. 
However, the selected projects are large, long-lived systems, mitigating the likelihood of bias in our report.
The replication of our study using diverse projects may prove useful. 
We report different metrics to cover all advantages and disadvantages of the methods. 
Also, we use SVM as the text classifier following the previous works; however, other classifiers may result in different classification performances. \textcolor{black}{We investigate the possibility of improvement in prediction performance using deep learning algorithms. The result is reported on the \href{https://github.com/HadiJahanshahi/SDABT}{GitHub page} of the paper.}

\section{Related Work}~\label{sec:related_works}
Several studies have been conducted to automate the bug triage process and decrease the cost of manual bug triage.
Different techniques have been employed to address this problem, e.g., text categorization, tossing graph, fuzzy set-based automatic bug triaging, role analysis-based automatic bug triage, information retrieval, and deep learning techniques.

Automatic bug triage using text classification is proposed by \citet{alenezi2013efficient,anvik2006should,xuan2017automatic} wherein they trained different classifiers such as SVM, Naive Bayes, and C4.5 on the history of bug fixes.
\citet{xuan2017automatic} proposed improvements for the Naive Bayes classifier by utilizing expectation-maximization based on the combination of labeled and unlabeled bug reports. 
They trained a classifier with a fraction of labeled bug reports. 
Then, they iteratively classified the unlabeled reports and fitted a new classifier with the labels of all the bug reports. 
These methods only aim to optimize the accuracy, ignoring many other aspects of the bug triaging process.

\citet{park2011costriage} proposed a method to optimize not only the accuracy but also the cost. 
This method combines the CBR model with a collaborative filtering recommender (CF) model, enhancing the recommendation quality of either approach alone.
\citet{alenezi2013efficient} presented a text mining approach to reduce the time and cost of bug triaging. 
They examined the use of four-term selection methods, namely, log odds ratio, chi-square, term frequency relevance frequency, and mutual information on the accuracy of bug assignment. 
They aimed to choose the most discriminating terms that describe bug reports. 
They then built the classifier using the Naive Bayes classifier on bug reports. 
They also incorporated cost to re-balance the load between developers considering their experience. 
\citet{kashiwa2020} also emphasized the distribution of the loads among developers. 
Their method aims to increase the number of bugs fixed by the next release. 
They formulated the bug triaging process as a multiple knapsack problem that maximizes the developers' preferences given a time limit. 
Consequently, it mitigates the task concentration to particular, experienced developers.
% \textcolor{black}{\citet{jahanshahi2021dabt} enhanced their work by limiting the feasible solution of their model using dependencies among bugs. They also emphasized the importance of using the bug-fixing time in the model objective. They showed that their proposed model reduces the total fixing time by these changes to the model.}

\citet{lee2020improving} presented a method that addresses the issues related to LDA fixing time calculation using a multiple LDA-based topic set. 
Their method improved the existing models by building two additional topic sets, partial topic set (PTS) and feature topic set (FTS). 
They showed that improved LDA has better classification accuracy. 
Also, \citet{xia2016improving} recommended a bug triaging method enhanced by specialized topic modeling, named multi-feature topic model (MTM), which extends LDA by considering bugs' components and products. 
Their proposed approach, TopicMiner, considers the topic distribution of a new bug report to make recommendations based on a developer's affinity to the topics and the feature combination.
Although these works rely on textual information, they do not consider developer engagement. 
\citet{ge2020high} proposed a method that overcomes this drawback. They build a high-quality dataset by combining the feature selection and instance selection and studied the impact of developer engagement on bug triage. 
They considered the product information along with the textual information in the bug report to recommend the best developer for a new bug report.

Many recent studies investigated automating the bug triage using deep learning techniques. 
\citet{lee2017applying} suggested using a Convolutional Neural Network (CNN) and word embedding to build an automatic bug triage. 
\citet{mani2019deeptriage} utilized an attention-based deep bi-directional RNN model (DBRNN-A) to automate bug triage. 
Their approach enables the model to learn the context representation over a long word sequence, as in a bug report. 
%Moreover, they compared their methods with four different classifiers, multinomial naive Bayes, cosine similarity-based classifier, support vector machines, and softmax (regression) based classifier. 
Their results show DBRNN-A, along with the softmax classifier, outperforms bag-of-words models.
\citet{guo2020developer} proposed a developer activity-based CNN method for bug triage that recommends a list of developers. 
They combined CNN with batch normalization and pooling to learn from the word vector representation of bug reports generated by Word2vec.

A significant characteristic of bug reports is their dependency. 
However, its importance is rarely acknowledged in the bug triage domain. 
%\citet{almhana2021considering} proposed an automated bugs triage to rank bug reports based on bug report dependency and priority. They defined the dependency between two bug reports as the number of common files to be inspected to localize the bugs. Afterwards, they used a multi-objective search algorithm to find the trade-off between bug priority and dependency to rank the bug reports when assigned to developers.
\citet{kumari2019quantitative} developed a bug dependency-based mathematical model to develop software reliability growth models. 
They interpreted the bug summary description and comments in terms of entropy that also measures the uncertainty and irregularity of the bug tracking system. 
In the bug triaging process, the incoming bugs are dynamic that makes the bug dependency graph uncertain. 
To address this issue, \citet{akbarinasaji2018partially} constructed a bug dependency graph considering two graph metrics, i.e., depth and degree. 
They proposed a Partially Observable Markov Decision Process model for sequential decision making to prioritize incoming bugs based on the bug fixing history and use Partially Observable Monte Carlo Planning to identify the best policies for prioritizing the bugs. %As exploring bug dependency is a time-consuming and challenging task, \citet{luaphol2018assembling} presented a method to cluster the relevant bug reports using a constrained k-means clustering algorithm. The first constraint added to the k-means algorithm is to provide the value of $k$ with the number of meta-bugs, and the second is to provide the meta-bug as the centroid of each cluster. Furthermore, they compared the three weighting methods TF, TF-IDF, and BM25 to find the most appropriate weighting method.

%In the reviews discussed above, \citet{anvik2006should,xuan2017automatic} used the bug report textual information to predict suitable developers, while \citet{park2011costriage} also considered reducing the bug fix time with predicting developers. Further, \citet{kashiwa2020,alenezi2013efficient} tried to balance the load among the developers. Some researchers studied the bug dependency \citet{almhana2021considering,kumari2019quantitative,akbarinasaji2018partially} to automate bug triage. Our proposed bug triaging method takes into account all the above-discussed aspects and assigns bugs to appropriate developers.

\textcolor{black}{Different from the previous studies, our proposed model, S-DABT, is a comprehensive model that captures the most important aspects of the bug triage task that are specified by the domain experts~\citep{akbarinasaji2018partially, kashiwa2020, park2011costriage}. 
It uses textual information to estimate the bug fixing time. 
Also, the information is fed into a classifier (e.g., SVM) to find the appropriate developers. 
However, instead of simply combining these values, S-DABT considers the importance of developers' available time slots. It uses developers' schedule and their ability to address multiple bugs at a time in its formulation.
Given that constraint, it also postpones the bugs blocked by others and cannot be solved currently, regardless of whether the blocking bug is an unassigned or assigned bug. 
Accordingly, it covers the objectives of the previous works subject to its novel constraints.}

\section{Concluding Remarks}~\label{sec:conclusion}
In this paper, \textcolor{black}{we proposed a schedule and dependency-aware bug triage method (S-DABT) that aims to reduce bug fixing time and infeasible assignment of blocked bugs while considering the developers' schedule and determining the exact assignment date.} 
S-DABT also takes into account the workloads of the developers in the bug triage process and alleviates task concentration on a small portion of developers. 
Accordingly, it reduces the number of overdue bugs before the next release. 

% S-DABT is enhanced by including additional constraints on the blocked bugs. 
Although S-DABT is primarily a triaging method, it also prioritizes the bugs such that both the complexity of the bug dependency graph and the total fixing time are reduced. \textcolor{black}{It extends DABT's awareness of bug dependencies by incorporating the constraint on blocked bugs that are assigned and yet to be fixed. 
This enhancement lowers the number of infeasible bug fixings due to bug dependencies in the long run.}
Through our experiments with four open-source software systems, S-DABT demonstrates a robust result in terms of the reduced overdue bugs, the improved fixing time of the assigned bugs, and the decreased complexity of the bug dependency graph. 
The model has lower accuracy compared to the other baselines. 
\textcolor{black}{However, it improves the accuracy of its predecessor, DABT. 
Through sensitivity analysis, we further demonstrate that it achieves higher assignment accuracy for different hyperparameter settings, which comes at the expense of increased average fixing time.}

\textcolor{black}{In this work, we assume that each developer may work on one or more bugs simultaneously. 
Unlike previous works that did not consider developers' different characteristics, this assumption is akin to the actual case where each developer has a different number of working days and different tendencies on being focused on single or multiple tasks.
A relevant venue for future research would be to formulate an IP model in which bug assignments to developers are restricted by the software release dates.
Accordingly, it minimizes the possibility of overdue bugs.
Another possible future research direction is to incorporate constraints that explicitly order the bug assignments in the IP model.
% ordering bug assignments within the model formula.
Currently, our model reorders bugs based on their importance (e.g., fixing time) as a post-processing step.
However, it can be incorporated in the model formulation, with the caveat that it may increase the complexity of the model.
}

\section*{Supplementary Materials}~\label{sec:supplementary}
To make the work reproducible, we publicly share our originally extracted dataset of one-decade bug reports, scripts, and analysis on \href{https://github.com/HadiJahanshahi/SDABT}{\textcolor{black}{GitHub}}.

{\singlespacing
\bibliographystyle{ACM-Reference-Format}
\bibliography{references}


\begin{thebibliography}{33}

%%% the Web address.

\ifx \showCODEN    \undefined \def \showCODEN     #1{\unskip}     \fi
\ifx \showDOI      \undefined \def \showDOI       #1{#1}\fi
\ifx \showISBNx    \undefined \def \showISBNx     #1{\unskip}     \fi
\ifx \showISBNxiii \undefined \def \showISBNxiii  #1{\unskip}     \fi
\ifx \showISSN     \undefined \def \showISSN      #1{\unskip}     \fi
\ifx \showLCCN     \undefined \def \showLCCN      #1{\unskip}     \fi
\ifx \shownote     \undefined \def \shownote      #1{#1}          \fi
\ifx \showarticletitle \undefined \def \showarticletitle #1{#1}   \fi
\ifx \showURL      \undefined \def \showURL       {\relax}        \fi
% invisible to TeX
\providecommand\bibfield[2]{#2}
\providecommand\bibinfo[2]{#2}
\providecommand\natexlab[1]{#1}
\providecommand\showeprint[2][]{arXiv:#2}

\bibitem[\protect\citeauthoryear{Akbarinasaji, Bener, and Neal}{Akbarinasaji
  et~al\mbox{.}}{2017}]%
        {LingeringbugsShirin2017}
\bibfield{author}{\bibinfo{person}{Shirin Akbarinasaji}, \bibinfo{person}{Ayse
  Bener}, {and} \bibinfo{person}{Adam Neal}.} \bibinfo{year}{2017}\natexlab{}.
\newblock \showarticletitle{A Heuristic for Estimating the Impact of Lingering
  Defects: Can Debt Analogy Be Used as a Metric?}. In
  \bibinfo{booktitle}{\emph{Proceedings of the 8th Workshop on Emerging Trends
  in Software Metrics}} \emph{(\bibinfo{series}{WETSoM '17})}.
  \bibinfo{publisher}{IEEE Press}, \bibinfo{address}{Buenos Aires, Argentina},
  \bibinfo{pages}{36–42}.
\newblock
\showISBNx{9781538628072}


\bibitem[\protect\citeauthoryear{Akbarinasaji, Kavaklioglu, Başar, and
  Neal}{Akbarinasaji et~al\mbox{.}}{2020}]%
        {akbarinasaji2018partially}
\bibfield{author}{\bibinfo{person}{Shirin Akbarinasaji}, \bibinfo{person}{Can
  Kavaklioglu}, \bibinfo{person}{Ayşe Başar}, {and} \bibinfo{person}{Adam
  Neal}.} \bibinfo{year}{2020}\natexlab{}.
\newblock \showarticletitle{Partially observable Markov decision process to
  generate policies in software defect management}.
\newblock \bibinfo{journal}{\emph{Journal of Systems and Software}}
  \bibinfo{volume}{163} (\bibinfo{year}{2020}), \bibinfo{pages}{110518}.
\newblock
\showISSN{0164-1212}


\bibitem[\protect\citeauthoryear{Alenezi, Magel, and Banitaan}{Alenezi
  et~al\mbox{.}}{2013}]%
        {alenezi2013efficient}
\bibfield{author}{\bibinfo{person}{Mamdouh Alenezi}, \bibinfo{person}{Kenneth
  Magel}, {and} \bibinfo{person}{Shadi Banitaan}.}
  \bibinfo{year}{2013}\natexlab{}.
\newblock \showarticletitle{Efficient Bug Triaging Using Text Mining.}
\newblock \bibinfo{journal}{\emph{JSW}} \bibinfo{volume}{8},
  \bibinfo{number}{9} (\bibinfo{year}{2013}), \bibinfo{pages}{2185--2190}.
\newblock


\bibitem[\protect\citeauthoryear{Anvik, Hiew, and Murphy}{Anvik
  et~al\mbox{.}}{2006}]%
        {anvik2006should}
\bibfield{author}{\bibinfo{person}{John Anvik}, \bibinfo{person}{Lyndon Hiew},
  {and} \bibinfo{person}{Gail~C. Murphy}.} \bibinfo{year}{2006}\natexlab{}.
\newblock \showarticletitle{Who Should Fix This Bug?}. In
  \bibinfo{booktitle}{\emph{Proceedings of the 28th International Conference on
  Software Engineering}} (Shanghai, China) \emph{(\bibinfo{series}{ICSE '06})}.
  \bibinfo{publisher}{Association for Computing Machinery},
  \bibinfo{address}{New York, NY, USA}, \bibinfo{pages}{361–370}.
\newblock
\showISBNx{1595933751}


\bibitem[\protect\citeauthoryear{{Bhattacharya} and {Neamtiu}}{{Bhattacharya}
  and {Neamtiu}}{2010}]%
        {Bhattacharya2010}
\bibfield{author}{\bibinfo{person}{P. {Bhattacharya}} {and} \bibinfo{person}{I.
  {Neamtiu}}.} \bibinfo{year}{2010}\natexlab{}.
\newblock \showarticletitle{Fine-grained incremental learning and multi-feature
  tossing graphs to improve bug triaging}. In \bibinfo{booktitle}{\emph{2010
  IEEE International Conference on Software Maintenance}}.
  \bibinfo{publisher}{IEEE}, \bibinfo{address}{Timi oara, Romania},
  \bibinfo{pages}{1--10}.
\newblock


\bibitem[\protect\citeauthoryear{Blei, Ng, and Jordan}{Blei
  et~al\mbox{.}}{2003}]%
        {blei2003lDA}
\bibfield{author}{\bibinfo{person}{David~M Blei}, \bibinfo{person}{Andrew~Y
  Ng}, {and} \bibinfo{person}{Michael~I Jordan}.}
  \bibinfo{year}{2003}\natexlab{}.
\newblock \showarticletitle{Latent dirichlet allocation}.
\newblock \bibinfo{journal}{\emph{the Journal of machine Learning research}}
  \bibinfo{volume}{3} (\bibinfo{year}{2003}), \bibinfo{pages}{993--1022}.
\newblock


\bibitem[\protect\citeauthoryear{Ge, Zheng, Wang, and Li}{Ge
  et~al\mbox{.}}{2020}]%
        {ge2020high}
\bibfield{author}{\bibinfo{person}{Xin Ge}, \bibinfo{person}{Shengjie Zheng},
  \bibinfo{person}{Jiahui Wang}, {and} \bibinfo{person}{Hui Li}.}
  \bibinfo{year}{2020}\natexlab{}.
\newblock \showarticletitle{High-Dimensional Hybrid Data Reduction for
  Effective Bug Triage}.
\newblock \bibinfo{journal}{\emph{Mathematical Problems in Engineering}}
  \bibinfo{volume}{2020} (\bibinfo{year}{2020}), \bibinfo{pages}{5102897}.
\newblock


\bibitem[\protect\citeauthoryear{Guo, Zhang, Yang, Chen, Guo, Li, and Li}{Guo
  et~al\mbox{.}}{2020}]%
        {guo2020developer}
\bibfield{author}{\bibinfo{person}{Shikai Guo}, \bibinfo{person}{Xinyi Zhang},
  \bibinfo{person}{Xi Yang}, \bibinfo{person}{Rong Chen}, \bibinfo{person}{Chen
  Guo}, \bibinfo{person}{Hui Li}, {and} \bibinfo{person}{Tingting Li}.}
  \bibinfo{year}{2020}\natexlab{}.
\newblock \showarticletitle{Developer activity motivated bug triaging: via
  convolutional neural network}.
\newblock \bibinfo{journal}{\emph{Neural Processing Letters}}
  \bibinfo{volume}{51}, \bibinfo{number}{3} (\bibinfo{year}{2020}),
  \bibinfo{pages}{2589--2606}.
\newblock


\bibitem[\protect\citeauthoryear{Gupta, Kumar, and Kapur}{Gupta
  et~al\mbox{.}}{2019}]%
        {Gupta2019}
\bibfield{author}{\bibinfo{person}{Viral Gupta}, \bibinfo{person}{Deepak
  Kumar}, {and} \bibinfo{person}{P.~K. Kapur}.}
  \bibinfo{year}{2019}\natexlab{}.
\newblock \showarticletitle{Optimizing the Defect Prioritization in Enterprise
  Application Integration}. In \bibinfo{booktitle}{\emph{Software
  Engineering}}, \bibfield{editor}{\bibinfo{person}{M.~N. Hoda},
  \bibinfo{person}{Naresh Chauhan}, \bibinfo{person}{S.~M.~K. Quadri}, {and}
  \bibinfo{person}{Praveen~Ranjan Srivastava}} (Eds.).
  \bibinfo{publisher}{Springer Singapore}, \bibinfo{address}{Singapore},
  \bibinfo{pages}{585--597}.
\newblock
\showISBNx{978-981-10-8848-3}


\bibitem[\protect\citeauthoryear{Jahanshahi, Cevik, Navas-Sú, Başar, and
  González-Torres}{Jahanshahi et~al\mbox{.}}{2022}]%
        {jahanshahi2020Wayback}
\bibfield{author}{\bibinfo{person}{Hadi Jahanshahi}, \bibinfo{person}{Mucahit
  Cevik}, \bibinfo{person}{José Navas-Sú}, \bibinfo{person}{Ayşe Başar},
  {and} \bibinfo{person}{Antonio González-Torres}.}
  \bibinfo{year}{2022}\natexlab{}.
\newblock \showarticletitle{Wayback Machine: A tool to capture the evolutionary
  behaviour of the bug reports and their triage process in open-source software
  systems}.
\newblock \bibinfo{journal}{\emph{Journal of Systems and Software}}
  \bibinfo{volume}{-} (\bibinfo{year}{2022}), \bibinfo{pages}{111308}.
\newblock
\showISSN{0164-1212}


\bibitem[\protect\citeauthoryear{Jahanshahi, Chhabra, Cevik, and
  Ba\c{s}ar}{Jahanshahi et~al\mbox{.}}{2021}]%
        {jahanshahi2021dabt}
\bibfield{author}{\bibinfo{person}{Hadi Jahanshahi}, \bibinfo{person}{Kritika
  Chhabra}, \bibinfo{person}{Mucahit Cevik}, {and} \bibinfo{person}{Ay\c{s}e
  Ba\c{s}ar}.} \bibinfo{year}{2021}\natexlab{}.
\newblock \showarticletitle{DABT: A Dependency-Aware Bug Triaging Method}. In
  \bibinfo{booktitle}{\emph{Evaluation and Assessment in Software Engineering}}
  (Trondheim, Norway) \emph{(\bibinfo{series}{EASE 2021})}.
  \bibinfo{publisher}{Association for Computing Machinery},
  \bibinfo{address}{New York, NY, USA}, \bibinfo{pages}{221–230}.
\newblock
\showISBNx{9781450390538}


\bibitem[\protect\citeauthoryear{Kashiwa and Ohira}{Kashiwa and Ohira}{2020}]%
        {kashiwa2020}
\bibfield{author}{\bibinfo{person}{Yutaro Kashiwa} {and} \bibinfo{person}{Masao
  Ohira}.} \bibinfo{year}{2020}\natexlab{}.
\newblock \showarticletitle{A Release-Aware Bug Triaging Method Considering
  Developers' Bug-Fixing Loads}.
\newblock \bibinfo{journal}{\emph{IEICE TRANSACTIONS on Information and
  Systems}} \bibinfo{volume}{103}, \bibinfo{number}{2} (\bibinfo{year}{2020}),
  \bibinfo{pages}{348--362}.
\newblock


\bibitem[\protect\citeauthoryear{Kumar and Yadav}{Kumar and Yadav}{2017}]%
        {Kumar2017}
\bibfield{author}{\bibinfo{person}{Chandan Kumar} {and}
  \bibinfo{person}{Dilip~Kumar Yadav}.} \bibinfo{year}{2017}\natexlab{}.
\newblock \showarticletitle{Software defects estimation using metrics of early
  phases of software development life cycle}.
\newblock \bibinfo{journal}{\emph{International Journal of System Assurance
  Engineering and Management}} \bibinfo{volume}{8}, \bibinfo{number}{4}
  (\bibinfo{year}{2017}), \bibinfo{pages}{2109--2117}.
\newblock


\bibitem[\protect\citeauthoryear{Kumari, Misra, Misra, Fernandez~Sanz,
  Damasevicius, and Singh}{Kumari et~al\mbox{.}}{2019}]%
        {kumari2019quantitative}
\bibfield{author}{\bibinfo{person}{Madhu Kumari}, \bibinfo{person}{Ananya
  Misra}, \bibinfo{person}{Sanjay Misra}, \bibinfo{person}{Luis
  Fernandez~Sanz}, \bibinfo{person}{Robertas Damasevicius}, {and}
  \bibinfo{person}{VB Singh}.} \bibinfo{year}{2019}\natexlab{}.
\newblock \showarticletitle{Quantitative quality evaluation of software
  products by considering summary and comments entropy of a reported bug}.
\newblock \bibinfo{journal}{\emph{Entropy}} \bibinfo{volume}{21},
  \bibinfo{number}{1} (\bibinfo{year}{2019}), \bibinfo{pages}{91}.
\newblock


\bibitem[\protect\citeauthoryear{Lee and Seo}{Lee and Seo}{2020}]%
        {lee2020improving}
\bibfield{author}{\bibinfo{person}{Dong-Gun Lee} {and}
  \bibinfo{person}{Yeong-Seok Seo}.} \bibinfo{year}{2020}\natexlab{}.
\newblock \showarticletitle{Improving bug report triage performance using
  artificial intelligence based document generation model}.
\newblock \bibinfo{journal}{\emph{Human-centric Computing and Information
  Sciences}} \bibinfo{volume}{10}, \bibinfo{number}{1} (\bibinfo{year}{2020}),
  \bibinfo{pages}{1--22}.
\newblock


\bibitem[\protect\citeauthoryear{Lee, Heo, Lee, Kim, and Jeong}{Lee
  et~al\mbox{.}}{2017}]%
        {lee2017applying}
\bibfield{author}{\bibinfo{person}{Sun-Ro Lee}, \bibinfo{person}{Min-Jae Heo},
  \bibinfo{person}{Chan-Gun Lee}, \bibinfo{person}{Milhan Kim}, {and}
  \bibinfo{person}{Gaeul Jeong}.} \bibinfo{year}{2017}\natexlab{}.
\newblock \showarticletitle{Applying Deep Learning Based Automatic Bug Triager
  to Industrial Projects}. In \bibinfo{booktitle}{\emph{Proceedings of the 2017
  11th Joint Meeting on Foundations of Software Engineering}} (Paderborn,
  Germany) \emph{(\bibinfo{series}{ESEC/FSE 2017})}.
  \bibinfo{publisher}{Association for Computing Machinery},
  \bibinfo{address}{New York, NY, USA}, \bibinfo{pages}{926–931}.
\newblock
\showISBNx{9781450351058}


\bibitem[\protect\citeauthoryear{Liu, Tian, Yu, Yang, Jia, Ma, and Xu}{Liu
  et~al\mbox{.}}{2016}]%
        {liu2016multi}
\bibfield{author}{\bibinfo{person}{Jin Liu}, \bibinfo{person}{Yiqiuzi Tian},
  \bibinfo{person}{Xiao Yu}, \bibinfo{person}{Zhijiang Yang},
  \bibinfo{person}{Xiangyang Jia}, \bibinfo{person}{Chuanxiang Ma}, {and}
  \bibinfo{person}{Zheng Xu}.} \bibinfo{year}{2016}\natexlab{}.
\newblock \showarticletitle{A multi-source approach for bug triage}.
\newblock \bibinfo{journal}{\emph{International Journal of Software Engineering
  and Knowledge Engineering}} \bibinfo{volume}{26}, \bibinfo{number}{09n10}
  (\bibinfo{year}{2016}), \bibinfo{pages}{1593--1604}.
\newblock


\bibitem[\protect\citeauthoryear{Mani, Sankaran, and Aralikatte}{Mani
  et~al\mbox{.}}{2019}]%
        {mani2019deeptriage}
\bibfield{author}{\bibinfo{person}{Senthil Mani}, \bibinfo{person}{Anush
  Sankaran}, {and} \bibinfo{person}{Rahul Aralikatte}.}
  \bibinfo{year}{2019}\natexlab{}.
\newblock \showarticletitle{DeepTriage: Exploring the Effectiveness of Deep
  Learning for Bug Triaging}. In \bibinfo{booktitle}{\emph{Proceedings of the
  ACM India Joint International Conference on Data Science and Management of
  Data}} (Kolkata, India) \emph{(\bibinfo{series}{CoDS-COMAD '19})}.
  \bibinfo{publisher}{Association for Computing Machinery},
  \bibinfo{address}{New York, NY, USA}, \bibinfo{pages}{171–179}.
\newblock
\showISBNx{9781450362078}


\bibitem[\protect\citeauthoryear{Park, Lee, Kim, Hwang, and Kim}{Park
  et~al\mbox{.}}{2011}]%
        {park2011costriage}
\bibfield{author}{\bibinfo{person}{Jin-woo Park}, \bibinfo{person}{Mu-Woong
  Lee}, \bibinfo{person}{Jinhan Kim}, \bibinfo{person}{Seung-won Hwang}, {and}
  \bibinfo{person}{Sunghun Kim}.} \bibinfo{year}{2011}\natexlab{}.
\newblock \showarticletitle{Cos\titlecap{Triage}: A Cost-Aware Triage Algorithm
  for Bug Reporting Systems}.
\newblock \bibinfo{journal}{\emph{Proceedings of the AAAI Conference on
  Artificial Intelligence}} \bibinfo{volume}{25}, \bibinfo{number}{1}
  (\bibinfo{date}{Aug.} \bibinfo{year}{2011}), \bibinfo{pages}{139--144}.
\newblock


\bibitem[\protect\citeauthoryear{Park, Lee, Kim, Hwang, and Kim}{Park
  et~al\mbox{.}}{2016}]%
        {park2016cost}
\bibfield{author}{\bibinfo{person}{Jin-woo Park}, \bibinfo{person}{Mu-Woong
  Lee}, \bibinfo{person}{Jinhan Kim}, \bibinfo{person}{Seung-won Hwang}, {and}
  \bibinfo{person}{Sunghun Kim}.} \bibinfo{year}{2016}\natexlab{}.
\newblock \showarticletitle{Cost-aware triage ranking algorithms for bug
  reporting systems}.
\newblock \bibinfo{journal}{\emph{Knowledge and Information Systems}}
  \bibinfo{volume}{48}, \bibinfo{number}{3} (\bibinfo{year}{2016}),
  \bibinfo{pages}{679--705}.
\newblock


\bibitem[\protect\citeauthoryear{Ren, Li, and Chen}{Ren et~al\mbox{.}}{2020}]%
        {Hao2020}
\bibfield{author}{\bibinfo{person}{Hao Ren}, \bibinfo{person}{Yanhui Li}, {and}
  \bibinfo{person}{Lin Chen}.} \bibinfo{year}{2020}\natexlab{}.
\newblock \showarticletitle{An Empirical Study on Critical Blocking Bugs}. In
  \bibinfo{booktitle}{\emph{Proceedings of the 28th International Conference on
  Program Comprehension}} (Seoul, Republic of Korea)
  \emph{(\bibinfo{series}{ICPC '20})}. \bibinfo{publisher}{Association for
  Computing Machinery}, \bibinfo{address}{New York, NY, USA},
  \bibinfo{pages}{72–82}.
\newblock
\showISBNx{9781450379588}


\bibitem[\protect\citeauthoryear{Sepahvand, Akbari, and Hashemi}{Sepahvand
  et~al\mbox{.}}{2020}]%
        {sepahvand2020}
\bibfield{author}{\bibinfo{person}{Reza Sepahvand}, \bibinfo{person}{Reza
  Akbari}, {and} \bibinfo{person}{Sattar Hashemi}.}
  \bibinfo{year}{2020}\natexlab{}.
\newblock \showarticletitle{Predicting the bug fixing time using word embedding
  and deep long short term memories}.
\newblock \bibinfo{journal}{\emph{IET Software}} \bibinfo{volume}{14},
  \bibinfo{number}{3} (\bibinfo{year}{2020}), \bibinfo{pages}{203--212}.
\newblock


\bibitem[\protect\citeauthoryear{Sung and Lee}{Sung and Lee}{2016}]%
        {sung2016optimal}
\bibfield{author}{\bibinfo{person}{Inkyung Sung} {and} \bibinfo{person}{Taesik
  Lee}.} \bibinfo{year}{2016}\natexlab{}.
\newblock \showarticletitle{Optimal allocation of emergency medical resources
  in a mass casualty incident: Patient prioritization by column generation}.
\newblock \bibinfo{journal}{\emph{European Journal of Operational Research}}
  \bibinfo{volume}{252}, \bibinfo{number}{2} (\bibinfo{year}{2016}),
  \bibinfo{pages}{623--634}.
\newblock


\bibitem[\protect\citeauthoryear{Tian, Lo, Xia, and Sun}{Tian
  et~al\mbox{.}}{2015}]%
        {tian2015}
\bibfield{author}{\bibinfo{person}{Yuan Tian}, \bibinfo{person}{David Lo},
  \bibinfo{person}{Xin Xia}, {and} \bibinfo{person}{Chengnian Sun}.}
  \bibinfo{year}{2015}\natexlab{}.
\newblock \showarticletitle{Automated prediction of bug report priority using
  multi-factor analysis}.
\newblock \bibinfo{journal}{\emph{Empirical Software Engineering}}
  \bibinfo{volume}{20}, \bibinfo{number}{5} (\bibinfo{year}{2015}),
  \bibinfo{pages}{1354--1383}.
\newblock


\bibitem[\protect\citeauthoryear{Valdivia-Garcia, Shihab, and
  Nagappan}{Valdivia-Garcia et~al\mbox{.}}{2018a}]%
        {ValdiviaGarcia2018}
\bibfield{author}{\bibinfo{person}{Harold Valdivia-Garcia},
  \bibinfo{person}{Emad Shihab}, {and} \bibinfo{person}{Meiyappan Nagappan}.}
  \bibinfo{year}{2018}\natexlab{a}.
\newblock \showarticletitle{Characterizing and predicting blocking bugs in open
  source projects}.
\newblock \bibinfo{journal}{\emph{Journal of Systems and Software}}
  \bibinfo{volume}{143} (\bibinfo{year}{2018}), \bibinfo{pages}{44--58}.
\newblock
\showISSN{0164-1212}


\bibitem[\protect\citeauthoryear{Valdivia-Garcia, Shihab, and
  Nagappan}{Valdivia-Garcia et~al\mbox{.}}{2018b}]%
        {Valdivia2018}
\bibfield{author}{\bibinfo{person}{Harold Valdivia-Garcia},
  \bibinfo{person}{Emad Shihab}, {and} \bibinfo{person}{Meiyappan Nagappan}.}
  \bibinfo{year}{2018}\natexlab{b}.
\newblock \showarticletitle{Characterizing and predicting blocking bugs in open
  source projects}.
\newblock \bibinfo{journal}{\emph{Journal of Systems and Software}}
  \bibinfo{volume}{143} (\bibinfo{year}{2018}), \bibinfo{pages}{44--58}.
\newblock
\showISSN{0164-1212}


\bibitem[\protect\citeauthoryear{Xia, Lo, Ding, Al-Kofahi, Nguyen, and
  Wang}{Xia et~al\mbox{.}}{2016}]%
        {xia2016improving}
\bibfield{author}{\bibinfo{person}{Xin Xia}, \bibinfo{person}{David Lo},
  \bibinfo{person}{Ying Ding}, \bibinfo{person}{Jafar~M Al-Kofahi},
  \bibinfo{person}{Tien~N Nguyen}, {and} \bibinfo{person}{Xinyu Wang}.}
  \bibinfo{year}{2016}\natexlab{}.
\newblock \showarticletitle{Improving automated bug triaging with specialized
  topic model}.
\newblock \bibinfo{journal}{\emph{IEEE Transactions on Software Engineering}}
  \bibinfo{volume}{43}, \bibinfo{number}{3} (\bibinfo{year}{2016}),
  \bibinfo{pages}{272--297}.
\newblock


\bibitem[\protect\citeauthoryear{Xuan, Jiang, Ren, Yan, and Luo}{Xuan
  et~al\mbox{.}}{2017}]%
        {xuan2017automatic}
\bibfield{author}{\bibinfo{person}{Jifeng Xuan}, \bibinfo{person}{He Jiang},
  \bibinfo{person}{Zhilei Ren}, \bibinfo{person}{Jun Yan}, {and}
  \bibinfo{person}{Zhongxuan Luo}.} \bibinfo{year}{2017}\natexlab{}.
\newblock \bibinfo{title}{Automatic Bug Triage using Semi-Supervised Text
  Classification}.
\newblock
\newblock
\showeprint[arxiv]{1704.04769}~[cs.SE]


\bibitem[\protect\citeauthoryear{Yang, Baek, Lee, and Lee}{Yang
  et~al\mbox{.}}{2017}]%
        {Yang2017}
\bibfield{author}{\bibinfo{person}{Geunseok Yang}, \bibinfo{person}{Seungsuk
  Baek}, \bibinfo{person}{Jung-Won Lee}, {and} \bibinfo{person}{Byungjeong
  Lee}.} \bibinfo{year}{2017}\natexlab{}.
\newblock \showarticletitle{Analyzing Emotion Words to Predict Severity of
  Software Bugs: A Case Study of Open Source Projects}. In
  \bibinfo{booktitle}{\emph{Proceedings of the Symposium on Applied Computing}}
  (Marrakech, Morocco) \emph{(\bibinfo{series}{SAC '17})}.
  \bibinfo{publisher}{Association for Computing Machinery},
  \bibinfo{address}{New York, NY, USA}, \bibinfo{pages}{1280–1287}.
\newblock
\showISBNx{9781450344869}


\bibitem[\protect\citeauthoryear{{Zaidi}, {Awan}, {Lee}, {Woo}, and
  {Lee}}{{Zaidi} et~al\mbox{.}}{2020}]%
        {Zaidi2020}
\bibfield{author}{\bibinfo{person}{S.~F.~A. {Zaidi}}, \bibinfo{person}{F.~M.
  {Awan}}, \bibinfo{person}{M. {Lee}}, \bibinfo{person}{H. {Woo}}, {and}
  \bibinfo{person}{C.~G. {Lee}}.} \bibinfo{year}{2020}\natexlab{}.
\newblock \showarticletitle{Applying Convolutional Neural Networks With
  Different Word Representation Techniques to Recommend Bug Fixers}.
\newblock \bibinfo{journal}{\emph{IEEE Access}}  \bibinfo{volume}{8}
  (\bibinfo{year}{2020}), \bibinfo{pages}{213729--213747}.
\newblock


\bibitem[\protect\citeauthoryear{Zaidi, Woo, and Lee}{Zaidi
  et~al\mbox{.}}{2022}]%
        {Zaidi2022}
\bibfield{author}{\bibinfo{person}{Syed Farhan~Alam Zaidi},
  \bibinfo{person}{Honguk Woo}, {and} \bibinfo{person}{Chan-Gun Lee}.}
  \bibinfo{year}{2022}\natexlab{}.
\newblock \showarticletitle{A Graph Convolution Network-Based Bug Triage System
  to Learn Heterogeneous Graph Representation of Bug Reports}.
\newblock \bibinfo{journal}{\emph{IEEE Access}}  \bibinfo{volume}{10}
  (\bibinfo{year}{2022}), \bibinfo{pages}{20677--20689}.
\newblock


\bibitem[\protect\citeauthoryear{Zhang, Gong, and Versteeg}{Zhang
  et~al\mbox{.}}{2013}]%
        {Zhang2013}
\bibfield{author}{\bibinfo{person}{Hongyu Zhang}, \bibinfo{person}{Liang Gong},
  {and} \bibinfo{person}{Steve Versteeg}.} \bibinfo{year}{2013}\natexlab{}.
\newblock \showarticletitle{Predicting bug-fixing time: An empirical study of
  commercial software projects}. In \bibinfo{booktitle}{\emph{2013 35th
  International Conference on Software Engineering (ICSE)}}.
  \bibinfo{publisher}{IEEE}, \bibinfo{address}{San Francisco CA USA},
  \bibinfo{pages}{1042--1051}.
\newblock


\bibitem[\protect\citeauthoryear{Zhang}{Zhang}{2020}]%
        {Zhang2020}
\bibfield{author}{\bibinfo{person}{Wei Zhang}.}
  \bibinfo{year}{2020}\natexlab{}.
\newblock \showarticletitle{Efficient Bug Triage For Industrial Environments}.
  In \bibinfo{booktitle}{\emph{2020 IEEE International Conference on Software
  Maintenance and Evolution (ICSME)}}. \bibinfo{publisher}{IEEE},
  \bibinfo{address}{Adelaide, Australia}, \bibinfo{pages}{727--735}.
\newblock


\end{thebibliography}
}

\end{document}